\documentclass[12pt]{iopart}
\usepackage{iopams}
\usepackage{hyperref} 
\usepackage{bm}
\usepackage{cite}

\usepackage{amsmath,amssymb,amsfonts}
\usepackage{graphicx,xcolor}
\usepackage{cancel}

\hypersetup{colorlinks=true, linktoc=all, linkcolor=blue}
\newcommand{\cbr}[1]{\left(#1\right)}
\newcommand{\sbr}[1]{\left[#1\right]}
\newcommand{\grad}{\nabla}

\newcommand{\dal}{\Box \phi}
\newcommand{\dpp}{\left(\nabla \phi \right)^2}

\renewcommand{\a}{\alpha}
\renewcommand{\b}{\beta}

\newcommand{\Hdcal}{ {}^d \! \mathcal{H} }

\newcommand{\Hd}{ {}^d \! H }

\newcommand{\Gdcal}{ {}^d \! \mathcal{G} }
\newcommand{\limitGcal}{ {}^3 \! \mathcal{G} }

\renewcommand{\d}[1]{\left(D-{#1}\right)}

\begin{document}

\title[4D Einstein-Gauss-Bonnet Gravity]{The 4D Einstein-Gauss-Bonnet Theory of Gravity: A Review}
\author{Pedro G. S. Fernandes$^{1a}$, Pedro Carrilho$^{2a,b}$,\\ Timothy Clifton$^{3a}$ and David J. Mulryne$^{4a}$}
\address{$^a$Department of Physics \& Astronomy, Queen Mary University of London, UK}
\address{$^b$Institute for Astronomy, The University of Edinburgh, UK}
\ead{$^1$p.g.s.fernandes@qmul.ac.uk, $^2$pedro.carrilho@ed.ac.uk, $^3$t.clifton@qmul.ac.uk, $^4$d.mulryne@qmul.ac.uk}

\begin{abstract}
We review the topic of 4D Einstein-Gauss-Bonnet gravity, which has been the subject of considerable interest over the past two years. Our review begins with a general introduction to Lovelock's theorem, and the subject of Gauss-Bonnet terms in the action for gravity. These areas are of fundamental importance for understanding modified theories of gravity, and inform our subsequent discussion of recent attempts to include the effects of a Gauss-Bonnet term in four space-time dimensions by re-scaling the appropriate coupling parameter. We discuss the mathematical complexities involved in implementing this idea, and review recent attempts at constructing well-defined, self-consistent theories that enact it. We then move on to consider the gravitational physics that results from these theories, in the context of black holes, cosmology, and weak-field gravity. We show that 4D Einstein-Gauss-Bonnet gravity exhibits a number of interesting phenomena in each of these areas.
\end{abstract}


\maketitle

\bibliographystyle{unsrt}

\tableofcontents

\newpage

\section{Introduction}
Ever since its inception there have been attempts to provide alternatives to Einstein's theory of General Relativity (GR). This began with the proposals of Weyl \cite{Weyl:1918ib} and Eddington \cite{eddington1921generalisation}, and has continued to the present day, where a plethora of different theories of gravity are now under active consideration in a number of different areas of physics \cite{Clifton:2011jh, Will:2014kxa}. These alternatives to GR are introduced for a variety of mathematical, philosophical and observational reasons, but almost all have the common function of generalizing the theory that Einstein initially proposed.

Holding a special place amongst this zoo of possibilities is the Einstein-Gauss-Bonnet theory, initially proposed by Lanczos \cite{lanczos1932elektromagnetismus, lanczos1938remarkable} and subsequently itself generalized by Lovelock \cite{lovelock1970divergence, Lovelock}. These theories are unique in requiring no extra fundamental fields beyond those that go into GR, while maintaining the property that the field equations of the theory can be written with no higher than second derivatives of the metric (a sufficient condition to prevent Ostrogradsky instability \cite{Ostrogradsky:1850fid}). They are therefore particularly well motivated, and hold a uniquely privileged position among the pantheon of alternatives to GR.

In this paper we review recent progress in determining the consequences of Einstein-Gauss-Bonnet gravity in four dimensional space-time. Although there is an obvious interest in studying gravity in four dimensions, the Einstein-Gauss-Bonnet extension of GR was for a long time thought to be trivial in this case. This changed in 2020, when Glavan \& Lin proposed a re-scaling of the coupling constant of the theory that potentially allowed for the consequences of Einstein-Gauss-Bonnet to be noticed even in the four-dimensional case \cite{Glavan:2019inb}. The theories that resulted from this idea have come to be known as ``4D Einstein-Gauss-Bonnet'' (4DEGB) gravity, and have a number of interesting properties.

In the sections that follow we will introduce the reader to the particular theories that fall under the umbrella of 4D Einstein-Gauss-Bonnet, as well as some of their most important properties and features. This will include a detailed discussion of the re-scaling proposed by Glavan \& Lin, as well as the necessary conditions required to implement it. We will then present and discuss the relevant physics that results from these theories in the cases of black holes, cosmology and weak gravitational fields. The 4D Einstein-Gauss-Bonnet theories display interesting phenomenology in each of the arenas, and we will summarize and discuss each of them as we progress. Before doing so, however, we will use the remainder of the present section to introduce some relevant concepts.

Throughout this review we will use a Lorentzian metric with signature $(-,+,+,+,\dots)$. Greek letters will be used to represent the space-time components of vectors and tensors in a coordinate basis, and Latin letters for spatial components. All expressions will be presented in units in which $G=c=1$, and we will use the notation $\square=\nabla_{\mu} \nabla^{\mu}$ where $\nabla_{\mu}$ represents the covariant derivative with respect to the metric $g_{\mu \nu}$.

\subsection{Lovelock Gravity}

Motivated by the idea of showing the uniqueness of Einstein's field equations, Lovelock asked which set of rank-2 tensors $A^{\mu \nu}$ could satisfy the following three conditions:
\begin{itemize}
\item[(i)] $A^{\mu \nu}= A^{\mu \nu} (g_{\rho \sigma}, g_{\rho \sigma, \tau} , g_{\rho \sigma, \tau \chi})$
\item[(ii)] $\nabla_{\nu} A^{\mu \nu} =0$
\item[(iii)] $A^{\mu \nu} = A^{ \nu \mu}$  .
\end{itemize}
Any such tensor would provide a plausible candidate for the left-hand side of the field equations of a geometric theory of gravity, and could be set as being proportional to the stress-energy tensor $T^{\mu \nu}$. Indeed, this seems to be a formalized version of the rationale that led Einstein to his formulation of the field equations in 1915, but here with the explicit aim of finding {\it all} possible field equations that would have consistent conservation and symmetry properties, as well as being free from Ostrogradski instabilities \cite{Ostrogradsky:1850fid}.

The question Lovelock posed had been partially answered much earlier by Weyl \cite{weyl1919space} and Cartan \cite{cartan1922equations}, who showed that if $A^{\mu \nu}$ is required to be linear in $g_{\rho \sigma, \tau \chi}$ then the only possibility is that $A^{\mu \nu}$ is a linear combination of the Einstein tensor and a cosmological constant term. By dropping the requirement of linearity, Lovelock found that there was a considerably broader class of solutions to the problem, each of which could serve as a suitable left-hand side in a geometric theory of gravity, without introducing any extra fundamental degrees of freedom beyond those that exist in the metric.

Lovelock's field equations can be derived from the following Lagrangian density:
\begin{equation}
\label{LoveL}
\mathcal{L} = \sqrt{-g} \sum_{j} \alpha_j \mathcal{R}^j \, ,
\end{equation}
where
\begin{equation} \hspace{-2cm}
\mathcal{R}^j \equiv \frac{1}{2^j} \delta^{\mu_1 \nu_1 \dots \mu_j \nu_j}_{\alpha_1 \beta_1 \dots \alpha_j \beta_j} \prod_{i=1}^{j} R^{\alpha_i \beta_i}_{\phantom{\alpha_i \beta_i} \mu_i \nu_i} \, , \qquad {\rm and} \qquad 
\delta^{\mu_1 \nu_1 \dots \mu_j \nu_j}_{\alpha_1 \beta_1 \dots \alpha_j \beta_j} \equiv j! \delta^{\mu_1}_{[\alpha_1} \delta^{\nu_1}_{\beta_1} \dots \delta^{\mu_j}_{\alpha_j} \delta^{\nu_j}_{\beta_j]} \, . \label{LLd}
\end{equation}
The $g$ in the Lagrangian density is the determinant of the metric, the $\alpha_j$ are a set of arbitrary constants, $R^{\mu}_{\phantom{\mu}\nu \rho \sigma}$ are the components of the Riemann tensor, and $\delta^{\mu}_{\nu}$ is the Kronecker delta. The square brackets denote anti-symmetrization in the usual way.

The tensor $A_{\mu \nu}$ that satisfies properties (i)-(iii) above can be generated from the Lagrangian density in Eq. (\ref{LoveL}) by integrating it over a region of $D$-dimensional space-time $\Omega$ to construct an action $S$, and then by varying with respect to the inverse metric $g^{\mu \nu}$. This gives
\begin{equation}
\delta S = \delta \int_{\Omega} d^Dx \, \mathcal{L} = \int_{\Omega} d^Dx \, \sqrt{-g} A_{\mu \nu} \, \delta g^{\mu \nu} + \int_{\partial \Omega} d^{D-1}x \, \sqrt{h} B \, ,
\end{equation}
where $h$ is the determinant of the induced metric on $\partial \Omega$, and where
\begin{equation} \label{LLA}
A^{\mu}_{\phantom{\mu} \nu} = - \sum_{j} \frac{\alpha_j}{2^{j+1}} \delta^{\mu \, \rho_1 \sigma_1 \dots \rho_j \sigma_j}_{\nu \, \alpha_1 \beta_1 \dots \alpha_j \beta_j} \prod_{i=1}^{j} R^{\alpha_i \beta_i}_{\phantom{\alpha_i \beta_i} \rho_i \sigma_i} \, .
\end{equation}
For further discussion of the derivation of this result, and an expression for the boundary term $B$, the reader is referred to the original literature \cite{lovelock1970divergence, Lovelock} and to the review \cite{Padmanabhan:2013xyr}.

\subsection{The Gauss-Bonnet Term}

One can immediately see that the sum in Eq. (\ref{LLA}) will terminate as soon as $2 j +1 > D$, where $D$ is the dimensionality of space-time. This follows from the definition of $\delta^{\mu_1 \nu_1 \dots \mu_j \nu_j}_{\alpha_1 \beta_1 \dots \alpha_j \beta_j}$ in Eq. (\ref{LLd}), as the number of possible values for each index must be greater than the number of lower indices in order for the quantity to be non-zero (otherwise at least two indices would have to take the same value, which would mean that it would vanish on anti-symmetrization). For even dimensional space-times we therefore have $D/2$ possible terms appearing in the tensor $A^{\mu}_{\phantom{\mu} \nu}$, while for odd dimensional space-times we have $(D+1)/2$ possible terms.

This means that in dimensions $D=1$ or $2$ there is only one term possible in the Lovelock tensor $A^{\mu}_{\phantom{\mu} \nu}$, and that this term will be of the functional form $\sim ({\it Riemann})^0$ (i.e. a constant). In dimensions $D=3$ and $4$ there are two possible terms, corresponding to a constant and to a term of the form $\sim({\it Riemann})^1$. In fact, this latter term is exactly the Einstein tensor so that in $D=4$
\begin{equation}
A^{\mu}_{\phantom{\mu} \nu}=-\frac{1}{2} \alpha_0 \delta^{\mu}_{\nu} + \alpha_1 \left( R^{\mu}_{\nu} - \frac{1}{2} \delta^{\mu}_{\nu} R \right) \, ,
\end{equation}
where $R_{\mu \nu}$ and $R$ are the Ricci curvature tensor and scalar, respectively. This is clearly the left-hand side of Einstein's equations, with the constants expressed in a slightly less familiar form. As a corollary of Lovelock's approach to gravity, we therefore have that Einstein's equations are the unique set of field equations that satisfy conditions (i)-(iii) above, which extends the result found by Weyl and Cartan to cases where $A^{\mu}_{\phantom{\mu} \nu}$ is allowed to be non-linear in second derivatives of the metric.

Continuing to higher dimensions, it can be shown that when $D>4$ Einstein's equations are {\it not} the most general set of field equations that obeys conditions (i)-(iii). In particular, in the case $D=5$ or $6$ the tensor $A^{\mu}_{\phantom{\mu} \nu}$ can contain {\it three} terms, with the last being order $\sim({\it Riemann})^2$ (i.e corresponding to $j=2$ in the sum in Eqs. (\ref{LoveL}) and (\ref{LLA})). This gives the Lagrangian density
\begin{equation} \label{LGauss-Bonnet}
\mathcal{L} = \sqrt{-g} \left[ \alpha_0 + \alpha_1 R + \alpha_2 \mathcal{G} \right] \, ,
\end{equation}
where
\begin{equation}
\mathcal{G} =  R^2 - 4 R_{\mu \nu} R^{\mu \nu} + R_{\mu \nu \rho \sigma} R^{\mu \nu \rho \sigma} 
\end{equation}
is known as the {\it Gauss-Bonnet} term. Extremization of the action associated with this Lagrangian gives the Lanczos tensor \cite{lanczos1932elektromagnetismus, lanczos1938remarkable}:
\begin{eqnarray} \hspace{-2cm} \label{AGauss-Bonnet}
A^{\mu}_{\phantom{\mu} \nu}
=-\frac{1}{2} \alpha_0 \delta^{\mu}_{\nu} &+ \alpha_1 \left( R^{\mu}_{\nu} - \frac{1}{2} \delta^{\mu}_{\nu} R \right) \\&+\alpha_2 \left( 2 R^{\mu \alpha \rho \sigma} R_{\nu \alpha \rho \sigma} - 4 R^{\rho \sigma} R^{\mu}_{\phantom{\mu} \rho \nu \sigma} - 4 R^{\mu \rho}R_{\nu \rho} +2 R \, R^{\mu}_{\nu} -\frac{1}{2} \delta^{\mu}_{\nu} \mathcal{G} \right) \, . \nonumber
\end{eqnarray}
This tensor provides an alternative set of field equations from those of Einstein, which has no higher than second derivatives of the metric, and which obeys the required symmetry and conservation properties in order for it to be set as being proportional to the stress-energy tensor $T^{\mu}_{\phantom{\mu} \nu}$.

The additional terms in the second line of Eq. (\ref{AGauss-Bonnet}) can be seen to vanish identically in $D=4$ and lower. This follows from the discussion above, and can also be understood as resulting from applying dimensionally-dependent identities to $A^{\mu}_{\phantom{\mu} \nu}$ \cite{lovelock1970dimensionally}. However, the same result can also be seen to be a consequence of the Chern theorem applied to the action that results from integrating Eq. (\ref{LGauss-Bonnet}) \cite{ChernTheorem} over the space-time manifold. In this latter case the integral of the Gauss-Bonnet term is equal to a constant with a value that depends on the Euler characteristic of the manifold, and which upon extremization contributes precisely zero to $A^{\mu}_{\phantom{\mu} \nu}$. It is for this reason that the Gauss-Bonnet term in $D=4$ is often referred to as a ``topological term'', and neglected. This is despite the fact that generically $\mathcal{G} \neq 0$ in $D=4$.

In dimensions $D>6$ there are further terms available in the Lovelock's theory, with the order of non-trivial new terms in powers of the Riemann tensor increasing consistently as the dimensionality of space-time increases. We will not consider these further possible terms here, but rather restrict ourselves to the Lagrangian that contains the Gauss-Bonnet term (\ref{LGauss-Bonnet}). Besides being the unique quadratic curvature combination appearing in the Lovelock Lagrangian, Gauss-Bonnet terms are of wide theoretical interest, as we will now describe. 

\subsection{Einstein-Gauss-Bonnet Gravity}

The combination of Einstein-Hilbert and Gauss-Bonnet terms in the gravitational action result in theories that have come to be known as Einstein-Gauss-Bonnet gravity. Such theories are of interest partly because string theory predicts that at the classical level Einstein's equations are subject to next-to-leading-order corrections that are typically described by higher-order curvature terms in the action. As we have just seen, the Gauss-Bonnet term is the unique term that is quadratic in the curvature and that results in second-order field equations.

As an example of how Einstein-Gauss-Bonnet gravity arises, it can be shown that M-theory compactified on a Calabi-Yau three-fold down to $D=5$ takes the effective form \cite{Ferrara:1996hh,Antoniadis:1997eg}
\begin{equation} \label{5DEinstein-Gauss-Bonnet}
S_{eff} = \int d^5x \sqrt{-g}\left(R + \frac{1}{16}c_2^{(I)}V_I\, \mathcal{G} \right),
\end{equation}
where $c_2^{(I)}V_I$ depends on the details of the Calabi-Yau manifold. This is nothing but the five-dimensional Lovelock theory presented in Eq. (\ref{LGauss-Bonnet}) (with suitably chosen $\alpha_i$). Gauss-Bonnet terms also occur in heterotic string theory \cite{Zwiebach:1985uq,Nepomechie:1985us,Callan:1986jb,Candelas:1985en,Gross:1986mw}, where the 1-loop effective action in the Einstein frame displays couplings of the form $\alpha^\prime e^{\phi} \mathcal{G}$ in the four-dimensional theory (where $\phi$ is a dynamical scalar field: the dilaton).

The mathematical foundations of theories containing both Einstein and Gauss-Bonnet terms have been extensively studied, with Choquet-Bruhat herself addressing the associated Cauchy problem \cite{choquet1988cauchy}, and the Hamiltonian problem being presented in Ref. \cite{teitelboim1987dimensionally}. Cosmological models have been particularly well studied in these theories, including during inflation, and in the context of ``brane'' cosmology (see Ref. \cite{deruelle2003quasi} for a review). They have also found application in the study of black hole thermodynamics and in the emergent gravity paradigm (see Ref. \cite{Padmanabhan:2013xyr} for a review). The reader will note, however, that the dimensionality of the manifold over which the integration is performed in the action in Eq. (\ref{5DEinstein-Gauss-Bonnet}) must necessarily be $D>4$, in order for there to be non-vanishing contributions to the field equations.

The appearance of Gauss-Bonnet terms in string-inspired theories of gravity has also motivated the consideration of four-dimensional theories of the form
\begin{equation} \label{dEinstein-Gauss-Bonnet}
S = \int d^4x \sqrt{-g} \left(R - \frac{1}{2} \dpp + f(\phi) \mathcal{G} \right) \, ,
\end{equation}
where here a kinetic term has been added for the scalar field. These scalar-tensor theories belong to the recently revived Horndeski class \cite{Horndeski} (see Ref. \cite{Kobayashi:2019hrl} for a review), and are related to generalized Galileon theories of gravity \cite{VanAcoleyen:2011mj,Charmousis:2015txa,Charmousis:2012dw}. In these theories it is possible to work in $D=4$ space-time dimensions, and still have non-vanishing contributions of the Gauss-Bonnet term to the field equations, due to the existence of the dilatonic scalar field $\phi$. We remark that when it comes to purely geometric terms, only couplings of the scalar field to the Ricci scalar and the Gauss-Bonnet term are allowed by Horndeski's theory.

These scalar-tensor variants of Einstein-Gauss-Bonnet theory are also well studied, and have been found to exhibit a rich phenomenology \cite{Sotiriou:2013qea, Sotiriou:2014pfa, Saravani:2019xwx, Delgado:2020rev,Doneva:2017bvd,Silva:2017uqg,Antoniou:2017acq,Cunha:2019dwb,Collodel:2019kkx,Dima:2020yac,Herdeiro:2020wei,Berti:2020kgk,Kanti:1995vq,Kleihaus:2011tg,Kleihaus:2015aje,Cunha:2016wzk,Blazquez-Salcedo:2017txk,Nojiri:2005vv,Jiang:2013gza,Kanti:2015pda,Chakraborty:2018scm,Odintsov:2018zhw,Odintsov:2019clh,Odintsov:2020zkl,Kanti:1998jd}. In particular, they are expected to produce viable models of inflation in the early universe, display spontaneous scalarization in compact objects, and admit novel black hole solutions that evade the no-hair theorems (see Ref. \cite{Herdeiro:2015waa} for a review).

\subsection{4D Einstein-Gauss-Bonnet Gravity}

In order to circumvent the stringent requirements of Lovelock's theory, and in an attempt to introduce the Gauss-Bonnet term in 4D gravity directly, Glavan \& Lin proposed rescaling the coupling constant $\alpha_2$ such that \cite{Glavan:2019inb}
\begin{equation} \label{GL}
\alpha_2 \rightarrow \frac{\alpha_2}{(D-4)} \, .
\end{equation}
This quantity is clearly divergent in the limit $D\rightarrow 4$, but Glavan \& Lin made the non-trivial suggestion that if this re-scaling were introduced into the Lanczos tensor (\ref{AGauss-Bonnet}) then the terms that contain this quantity as a factor might remain finite and non-zero. That is, they postulated that the divergence they introduced into $\alpha_2$ might be sufficient to cancel out the fact that additional terms in Eq. (\ref{AGauss-Bonnet}) tend to zero as $D \rightarrow 4$. If this were the case, then the Gauss-Bonnet term would be allowed to have a direct effect in the 4D theory of gravity.

Motivation for this radical new approach came from the trace of the Lanczos tensor (\ref{AGauss-Bonnet}), which in $D$ dimensions gives
\begin{equation} \label{traceA}
A^{\mu}_{\phantom{\mu} \mu} = -\frac{1}{2} D \alpha_0 - \frac{1}{2} (D-2) \alpha_1 R  - \frac{1}{2} (D-4) \alpha_2 \mathcal{G} \, .
\end{equation}
The vanishing of the term from the Einstein tensor in $D=2$ and the vanishing of the Gauss-Bonnet term in $D=4$ are both made explicit here, and both can be seen to be due to a pre-factor of the form $(D-n)$ (recall that $R$ and $\mathcal{G}$ can be non-zero only if $D>1$ and $D>3$, respectively). Using the re-scaling in Eq. (\ref{GL}) can then be seen to entirely remove the factor that usually results in the contribution from the Gauss-Bonnet term vanishing, and leaves a term that can in general be non-zero in the limit $D\rightarrow 4$. 

The additional term that results in the trace of the field equations (\ref{traceA}), after the re-scaling given in Eq. (\ref{GL}), are strongly motivated from studying quantum corrections to the stress-energy tensor in the presence of gravity. In this case the renormalized vacuum expectation value for the trace of $T_{\mu \nu}$ includes terms that are proportional to $\mathcal{G}$ \cite{duff1994twenty}, in just the same way that they are found in the trace of the left-hand side of the field equations in Eq. (\ref{traceA}). This is known as the ``conformal'' or ``trace'' anomaly in the quantum field theory literature, and a natural interpretation of the Glavan \& Lin re-scaling is that it is a way of accounting for the conformal anomaly in the gravitational sector of the theory. The reader may also note that a similar procedure to the re-scaling (\ref{GL}) has also been successfully applied to the Einstein term in the limit $D \rightarrow 2$ \cite{Mann:1992ar}, in order to remove the factor of $(D-2)$ that would otherwise result from Einstein's equations being entirely absent. We will return to this particular point later on.

It is the re-scaling presented in Eq. (\ref{GL}), and the ideas, phenomenology and theories that have resulted from it, that are the subject of this review. There has been a flurry of activity surrounding this idea in the year since it was published in {\it Physical Review Letters} \cite{Konoplya:2020bxa,Guo:2020zmf,Fernandes:2020rpa,Wei:2020ght,Konoplya:2020qqh,Hegde:2020xlv,Casalino:2020kbt,Ghosh:2020vpc,Doneva:2020ped,Zhang:2020qew,Ghosh:2020syx,Konoplya:2020ibi,Konoplya:2020juj,Kumar:2020owy,Kumar:2020uyz,Zhang:2020qam,HosseiniMansoori:2020yfj,Wei:2020poh,Singh:2020nwo,Churilova:2020aca,Islam:2020xmy,Mishra:2020gce,Kumar:2020xvu,Nojiri:2020tph,Singh:2020xju,Li:2020tlo,Heydari-Fard:2020sib,Konoplya:2020cbv,Jin:2020emq,Liu:2020vkh,Zhang:2020sjh,EslamPanah:2020hoj,NaveenaKumara:2020rmi,Aragon:2020qdc,Malafarina:2020pvl,Yang:2020czk,Cuyubamba:2020moe,Ying:2020bch,Shu:2020cjw,Casalino:2020pyv,Rayimbaev:2020lmz,Liu:2020evp,Zeng:2020dco,Ge:2020tid,Jusufi:2020yus,Churilova:2020mif,Kumar:2020sag,Alkac:2020zhg,Ghosh:2020cob,Yang:2020jno,Liu:2020yhu,Devi:2020uac,Jusufi:2020qyw,Konoplya:2020der,Qiao:2020hkx,Liu:2020evp,Samart:2020sxj,Banerjee:2020stc,Narain:2020qhh,Dadhich:2020ukj,Chakraborty:2020ifg,Singh:2020mty,Banerjee:2020yhu,Narain:2020tsw,Haghani:2020ynl,Lin:2020kqe,Shaymatov:2020yte,MohseniSadjadi:2020jmc,Banerjee:2020dad,Svarc:2020fia,Hegde:2020yrd,Li:2020vpo,Wang:2020pmb,Gao:2020vhw,Zhang:2020khz,Jafarzade:2020ilt,Ghaffarnejad:2020cru,Jafarzade:2020ova,Farsam:2020pfl,Colleaux:2020wfv,Mu:2020szg,Donmez:2020rnf,Hansraj:2020rvc,Junior:2020gnu,Abdujabbarov:2020jla,MohseniSadjadi:2020jmc,Li:2020spm,Zhang:2020obn,Li:2020ozr,Lin:2021noq,Zahid:2021vdy,Kruglov:2021pdp,Liu:2021zmi,Liu:2021fzr,Zhang:2021raw,Meng:2021huz,Ding:2021iwv,Babar:2021exh,Wu:2021zyl,Donmez:2021fbk,Chen:2021gwy,Feng:2020duo,Garcia-Aspeitia:2020uwq,Wang:2021kuw,Motta:2021hvl,Kruglov:2021stm,Li:2021izh,Atamurotov:2021imh,Bousder:2021yjr,Ovgun:2021ttv, Gurses:2020ofy,Gurses:2020rxb,Arrechea:2020evj,Arrechea:2020gjw,Bonifacio:2020vbk,Ai:2020peo,Mahapatra:2020rds,Hohmann:2020cor,Cao:2021nng,Heydari-Fard:2021ljh, Kruglov:2021btd,Ghorai:2021uby,Ghaffarnejad:2021zbx,Zhang:2021kha,Mishra:2021qhi,Shah:2021rob,Eslamzadeh:2021rvx,Marks:2021fpe,Bousder:2021ofq,Kruglov:2021rqf,Pretel:2021czp,Kong:2021qiu,Y:2021ybx,Vieira:2021doo,Gyulchev:2021dvt,Badia:2021kpk,Tangphati:2021wng,Huang:2021bdm,Singh:2021iwv,Dimov:2021fbm,Li:2021ohz,Li:2021wqa,Chatterjee:2021ops,Singh:2021xbk, Lu:2020iav,Kobayashi:2020wqy,Mann:1992ar,Fernandes:2020nbq,Hennigar:2020lsl,Fernandes:2021dsb,Aoki:2020lig,Shahidi:2021rnu, Clifton:2020xhc,Charmousis:2021npl,Tian:2020nzb,Hennigar:2020fkv,Lu:2020mjp,Ma:2020ufk,Gabadadze:2020tvt,Easson:2020mpq,Hennigar:2020drx,Wang:2020uiq,Easson:2020bgk,Fernandes:2021ysi,Aoki:2020iwm,Aoki:2020ila,Yao:2020tur,Pan:2021jii,Heydari-Fard:2021qdc,Banerjee:2021bmv,Narzilloev:2021jtg,Sengupta:2021mpf}, and we aim to bring some of this activity together into a single source, so that the interested reader can use it as a guide to the work that has been performed, and a source of references for further reading. In particular, we aim to present a balanced guide to the ways in which the proposed re-scaling (\ref{GL}) can be considered a viable method of introducing the consequences of a Gauss-Bonnet term in 4D, as well as those in which it cannot. We will draw on the work of many authors for this presentation, who will be referenced as we proceed.

\section{Einstein-Gauss-Bonnet Theory in 4D}\label{sec:4D EGB}

In this section we will discuss in more detail the proposal of Glavan \& Lin to re-scale the coupling constant of the Gauss-Bonnet term \cite{Glavan:2019inb}, before discussing the concerns and criticisms that have been raised about this idea, and the theories that have resulted from it.

\subsection{Glavan \& Lin theory and solutions}
\par The presentation here closely follows that of Ref. \cite{Glavan:2019inb}. Let us start by considering the typical Einstein-Gauss-Bonnet action, where for the moment we neglect any contributions from matter fields, focusing on the purely gravitational sector
\begin{equation}
S = \frac{1}{16\pi G}\int d^Dx \sqrt{-g} \left(-2\Lambda + R + \hat \alpha \mathcal{G}\right) \, ,
\label{eq:actionEinstein-Gauss-Bonnet}
\end{equation}
where $\hat \alpha$ is a constant. The reader will note that the number of space-time dimensions $D$ is not yet specified. Varying and extremizing the action with respect to the metric results in the field equations of the theory, which read
\begin{equation}
G_{\mu \nu} + \Lambda\, g_{\mu \nu} = \hat \alpha\,  H_{\mu \nu} \, ,
\label{eq:feqsGauss-Bonnet}
\end{equation}
where
\begin{equation}
H_{\mu \nu} = 15 \delta_{\mu [ \nu} R^{\rho \sigma}_{\phantom{\rho \sigma} \rho \sigma} R^{\alpha \beta}_{\phantom{\alpha \beta} \alpha \beta]} = -2 \left(R R_{\mu \nu} -2 R_{\mu \alpha \nu \beta} R^{\alpha \beta} + R_{\mu \alpha \beta \sigma}R_{\nu}^{\,\,\alpha \beta \sigma}-2R_{\mu \alpha}R_{\nu}^{\,\,\alpha}-\frac{1}{4}g_{\mu \nu} \mathcal{G} \right).
\label{eq:Gauss-Bonnettensor}
\end{equation}
The right-hand side of this equation is anti-symmetrized over five indices, and so must vanish in dimensions $D<5$. Up until this point, the presentation has been that of the usual Einstein-Gauss-Bonnet theory. The novelty added in Ref. \cite{Glavan:2019inb} is the possibility that the vanishing of $H_{\mu \nu}$ in four dimensions might be cancelled by re-scaling the coupling constant of the Gauss-Bonnet term, such that
\begin{equation}
\hat \alpha = \frac{\alpha}{D-4} \, ,
\label{eq:rescale}
\end{equation}
for some new \textit{finite} coupling constant $\alpha$ as we take the limit $D\to 4$. That this might be a viable possibility is suggested by the trace of the field equations \eqref{eq:feqsGauss-Bonnet}, which contains a contribution from the Gauss-Bonnet term that takes the form
\begin{equation}
g^{\mu \nu}H_{\mu \nu} = \frac{1}{2}(D-4)\, \mathcal{G} \, .
\end{equation}
It is clear that in this case the multiplicative factor of $(D-4)$ would be precisely cancelled by the suggested re-scaling of $\hat \alpha$, which would leave a non-vanishing contribution to the trace of the field equations as $D \to 4$: \footnote{Note that $\mathcal{G}$ itself is not required to vanish in the four-dimensional limit.}
\begin{equation}
\hat \alpha g^{\mu \nu}H_{\mu \nu} = \frac{\alpha}{\cancel{(D-4)}} \frac{1}{2}\cancel{(D-4)} \mathcal{G}= \frac{\alpha}{2}\mathcal{G}.
\end{equation}

\par The authors of Ref. \cite{Glavan:2019inb} suggest that this non-vanishing contribution may not be exclusive to the trace of the field equations, but could be manifest in the full theory. To support this claim they note that one can observe that the field equations written in differential form are
\begin{equation}
\varepsilon_{a_D} = \sum_{p=0}^{D/2-1}\alpha_p\left(D \!-\! 2p\right)\epsilon_{a_1...a_D} R^{a_1,a_2}\wedge...\wedge R^{a_{2p-1},a_{2p}} \wedge e^{a_{2p+1}}\wedge...\wedge e^{a_{D-1}}=0,
\label{eq:1storderform}
\end{equation}
where $e^a$ is the \textit{vielbein}. It may be noted here that the $(D-4)$ factor emerges in the field equations in this case, as $p=2$ for the Gauss-Bonnet contribution. While this is an intriguing suggestion, it appears that the desired result does not follow quite so straightforwardly, as we will discuss below. Nevertheless, there do exist $D$-dimensional space-times, which in the limit $D \rightarrow 4$ are well behaved under the proposed re-scaling.

\par Let us now focus on three important examples of physically interesting $D$-dimensional space-times: maximally-symmetric space-time, spherically-symmetric space-time and the homogeneous and isotropic FRW space-time. Starting with the maximally-symmetric space-time, we have
\begin{equation}
R_{\mu \alpha \nu \beta} = \frac{R}{D(D-1)} \left(g_{\mu \nu} g_{\alpha \beta}-g_{\mu \beta}g_{\alpha \nu} \right),
\end{equation}
with the Ricci scalar $R$ being constant. From this, one can prove that
\begin{equation}
H_{\mu \nu} = \frac{(D-4)(D-3)(D-2)}{2D^2(D-1)}g_{\mu \nu} R^2,
\end{equation}
thus resulting in the following non-trivial contribution to the field equations under the singular re-scaling of Eq. \eqref{eq:rescale}:
\begin{equation}
\lim_{D\to 4} \hat \alpha H_{\mu \nu} = \lim_{D\to 4}\frac{\alpha}{\cancel{(D-4)}} \frac{\cancel{(D-4)}(D-3)(D-2)}{2D^2(D-1)}g_{\mu \nu} R^2 = \frac{\alpha}{48} g_{\mu \nu} R^2.
\end{equation}
Under these conditions, there are two branches of solutions of the field equations \eqref{eq:feqsGauss-Bonnet} where the constant Ricci scalar acts as an effective cosmological constant, $\Lambda_{\rm eff}$, which obeys
\begin{equation}
\Lambda_{\rm eff}^\pm =  -\frac{6}{\alpha}\left(1\pm \sqrt{1+\frac{4\alpha \Lambda}{3}} \right).
\end{equation}
The existence of two branches of solutions is well-known in the higher-dimensional Einstein-Gauss-Bonnet theory (see e.g. \cite{BoulwareDeser}) and remains a feature of 4D EGB that will accompany us throughout this work. 

The two branches found above are fundamentally different. Assuming the Gauss-Bonnet contribution is a small correction to the theory, such that $\alpha \ll 1$, one obtains from the positive branch that
\begin{equation}
\Lambda^+_{\rm eff} \approx -\frac{12}{\alpha} -4\Lambda + \mathcal{O}\left(\alpha\right),
\end{equation}
while the negative branch gives
\begin{equation}
\Lambda^-_{\rm eff} \approx 4\Lambda + \mathcal{O}\left(\alpha\right).
\end{equation}
Clearly the positive branch does not possess a well-defined limit as $\alpha$ vanishes, whereas in the negative branch we recover the dynamics of GR. For this reason, the positive branch is dubbed the \textit{Gauss-Bonnet branch}, and the negative one the \textit{GR branch}.

\par At the level of perturbation theory, we perturb the metric as
\begin{equation}
g_{\mu \nu} = \overline{g}_{\mu \nu} + h_{\mu \nu},
\end{equation}
around the maximally symmetric space-time $\overline{g}_{\mu \nu}$. The linear perturbations in $D=4$ are then described by (see e.g. \cite{Fan:2016zfs} for details)
\begin{eqnarray}
\hspace{-2cm} \nonumber
 \biggl( 1+ \frac{4\alpha\Lambda}{3} \biggr)
\biggl[ \nabla^\rho \nabla^\mu h_{\nu \rho} + \nabla_\nu \nabla_\rho h^{\mu \rho}
		 - \square {h^\mu}_\nu -  \nabla^\mu \nabla_\nu {h^\rho}_\rho
	\\\hspace{2cm}+ \delta^{\mu}_{\nu} 
	\Bigl( \square {h^\rho}_\rho - \nabla_\rho \nabla_\sigma h^{\rho \sigma} \Bigr)
	+ \Lambda \Bigl( \delta^\mu_\nu {h^\rho}_\rho - 2 {h^\mu}_\nu \Bigr)
	\biggr] = 0 \, ,
\label{eq:linearized}
\end{eqnarray}
where the correction from the Gauss-Bonnet term can be observed to amount to an overall factor in the equation of motion, while the term in brackets is the same as in GR. Thus, just as in GR, the graviton has two degrees of freedom and the Gauss-Bonnet contribution to the linearized dynamics is trivial.

\par The same procedure can be performed for an FRW background
\begin{equation}
ds^2 = -dt^2 + a(t)^2 d\textbf{x}^2 \, ,
\end{equation}
where $a(t)$ is the scale factor. We define the Hubble rate as $H=\dot a/a$ and supplement the 4D EGB theory with matter in the form of a perfect fluid with stress-energy tensor $T^{\mu}_{\phantom{\mu}\nu} = \{-\rho, p,p,p,\ldots\}$, where $\rho$ and $p$ are the energy density and pressure of the matter fields. Under these conditions, the following set of (modified) Friedmann equations can be obtained
\begin{equation}
H^2+\alpha H^4 = \frac{8\pi G}{3} \rho + \frac{\Lambda}{3},
\end{equation}
\begin{equation}
\left(1+2\alpha H^2 \right) \dot H = -4\pi G (\rho+p),
\end{equation}
while the matter fields obey the standard continuity equation $\dot \rho + 3H\cbr{\rho+p}=0$. Interestingly, these Friedmann equations have exactly the same form as the ones obtained in holographic cosmology \cite{Apostolopoulos:2008ru,Bilic:2015uol}, from the generalized uncertainty principle \cite{Lidsey:2009xz}, by considering quantum entropic corrections \cite{Cai:2008ys}, and from gravity with a conformal anomaly \cite{Lidsey:2008zq}.

\par If we now consider transverse and traceless parts of the metric fluctuations, which describe gravitational waves, by perturbing as
\begin{equation}
g_{ij}=a^2(\delta_{ij}+\gamma_{ij}),
\end{equation}
where $\partial_i \gamma_{ij}=0$ and $\gamma_{ii}=0$, then we obtain in the four-dimensional limit a well-defined equation of motion, as the $(D-4)$ factors once again cancel. This gives
\begin{equation}
\ddot{\gamma}_{ij}+\cbr{3+\frac{4\alpha \dot H}{1+2\alpha H^2}}H \dot{\gamma}_{ij}-c_s^2\frac{\partial^2 \gamma_{ij}}{a^2}=0, \qquad {\rm where} \qquad c_s^2=1+\frac{4\alpha \dot H}{1+2\alpha H^2} \, .
\end{equation}
We observe that the Gauss-Bonnet contribution alters the Hubble friction and the sound speed, potentially leading to some non-trivial observational effects, which should be expected to be especially relevant in the early Universe.

\par Employing a general static and spherically-symmetric line-element, of the form
\begin{equation}
ds^2 = -f(r) e^{2\delta(r)} dt^2 + \frac{dr^2}{f(r)} + r^2 d\Omega_{D-2} \,,
\end{equation}
the field equations reveal, once again, an overall $(D-4)$ factor, leading to the following solution in the limit $D \rightarrow 4$:
\begin{equation}
f(r) = 1+\frac{r^2}{2\alpha} \cbr{1\pm \sqrt{1+\frac{8GM\alpha}{r^3}}}, \qquad {\rm and} \qquad \delta(r) = 0 \, .
\label{eq:BHsolution}
\end{equation}
This solution is extremely interesting for several reasons. First, it is highly reminiscent of the Boulware-Deser black hole from the higher-dimensional Einstein-Gauss-Bonnet theory \cite{BoulwareDeser}, which reads
\begin{equation}
f(r) = 1+\frac{r^2}{2 \alpha(D-3)(D-4)} \cbr{1\pm \sqrt{1+\frac{8GM \alpha(D-3)(D-4)}{r^{D-1}}}}\, ,
\end{equation}
and which also has $\delta(r) = 0$.
Second, these 4D EGB black hole solutions are exactly the same as solutions that appear in other contexts, namely by considering gravity with a conformal anomaly \cite{Cai:2009ua,Cai:2014jea}, entropy corrections to the black hole entropy \cite{Cognola:2013fva} and more interestingly as a solution to the proposed UV completion of gravity, Ho\v{r}ava-Lifshitz gravity \cite{Horava:2009uw}, known as the Kehagias-Sfetsos spacetime \cite{Kehagias:2009is}.

\subsection{Concerns and shortcomings}

\par As alluded to above, the novel approach of Glavan \& Lin has been met with a healthy amount of skeptical scrutiny \cite{Gurses:2020ofy,Gurses:2020rxb,Arrechea:2020evj,Arrechea:2020gjw,Bonifacio:2020vbk,Ai:2020peo,Mahapatra:2020rds,Hohmann:2020cor,Cao:2021nng}. Here we discuss these criticisms, and present the arguments they involve.

Let us begin with Refs. \cite{Gurses:2020ofy,Gurses:2020rxb,Arrechea:2020evj,Arrechea:2020gjw}, which have shown that the tensor resulting from the variation of the Gauss-Bonnet term, $H_{\mu \nu}$ given in Eq. \eqref{eq:Gauss-Bonnettensor}, can be written in $D$ dimensions in terms of the Weyl tensor as
\begin{equation}
H_{\mu \nu} =2\cbr{ H^{(1)}_{\mu \nu} + H^{(2)}_{\mu \nu} } \, ,
\end{equation}
where
\begin{equation}
H^{(1)}_{\mu \nu} = C_{\mu \alpha \beta \sigma} C_{\nu}^{\,\, \alpha \beta \sigma}-\frac{1}{4}g_{\mu \nu}C_{\alpha \beta \sigma \rho } C^{\alpha \beta \sigma \rho}\, ,
\end{equation}
and
\begin{equation}
\begin{aligned}
H^{(2)}_{\mu \nu} = \frac{\d4 \d3}{\d2\d1} \bigg[& -\frac{2\d1}{\d3}C_{\mu \rho \nu \sigma} R^{\rho \sigma} - \frac{2\d1}{\d2} R_{\mu \rho}R_{\nu}^\rho + \frac{D}{\d2} R_{\mu \nu}R \\& \hspace{1cm}+\frac{1}{\d2} g_{\mu \nu} \cbr{ \d1 R_{\rho \sigma}R^{\rho \sigma} - \frac{D+2}{4}R^2 }\bigg] \, ,
\end{aligned}
\end{equation}
and where here the $D$-dimensional expression for the Weyl tensor should be understood to be taken as
\begin{equation}
C_{\mu \alpha \nu \beta} = R_{\mu \alpha \nu \beta} - \frac{2}{D-2} \cbr{g_{\mu [\nu}R_{\beta]\alpha} - g_{\alpha [\nu}R_{\beta]\mu}} + \frac{2}{\cbr{D-1}\cbr{D-2}} R g_{\mu[\nu} g_{\beta]\alpha}\, .
\end{equation}
Now, while it is the case that in the limit $D \rightarrow 4$ the term $\hat{\alpha} H^{(2)}_{\mu \nu}$ is well-defined, such that
\begin{equation}
\lim_{D\to 4} \frac{H^{(2)}_{\mu \nu}}{\d4} = -C_{\mu \rho \nu \sigma} R^{\rho \sigma} - \frac{1}{2} R_{\mu \rho}R_{\nu}^\rho+\frac{1}{3}R_{\mu \nu}R + \frac{1}{4}g_{\mu \nu} \cbr{ R_{\rho \sigma}R^{\rho \sigma} - \frac{1}{2} R^2 }\, ,
\end{equation}
the same limit of $\hat{\alpha} {H^{(1)}_{\mu \nu}}$ is not. That is because $H^{(1)}_{\mu \nu} = 0$ vanishes identically in four dimensions, as the Riemann tensor loses independent components as one lowers the space-time dimension (a result analogous to $G_{\mu \nu}=0$ in 2 dimensions, and as discussed in the introduction). The poor behaviour of $\hat{\alpha} {H^{(1)}_{\mu \nu}}$ in the 4-dimensional limit is problematic, but if one were to simply ignore the above contribution to the field equations, the finite part resulting from the Gauss-Bonnet term would not be covariantly conserved, 
which would clearly be unacceptable.

Revisiting the arguments that employ the first-order formalism outlined in Eq. \eqref{eq:1storderform}, Refs. \cite{Gurses:2020ofy,Gurses:2020rxb,Arrechea:2020evj,Arrechea:2020gjw} argue that one cannot simply re-scale the coupling constant and take the four-dimensional limit. To see why, one can re-cast Eq. \eqref{eq:1storderform} in terms of space-time indices and take the Hodge dual of the $\d1$-form obtaining (see Ref. \cite{Gurses:2020ofy} for details)
\begin{equation}
\varepsilon_{\nu \alpha} = 2\d4! H_{\nu \alpha}=0,
\end{equation}
where $H_{\nu \alpha}$ is the Gauss-Bonnet tensor defined in Eq. \eqref{eq:Gauss-Bonnettensor}. The pre-factor here no longer vanishes when $D =4$, and the result is therefore that this approach does not lead to a well-defined $\d4$ factor in front of the field equations in the metric formulation. Again, this is obviously problematic for the proposed re-scaling procedure.

\par At the level of perturbation theory the Glavan \& Lin approach also seems ill-defined. Although at first-order there are no divergences, as observed in Eq. \eqref{eq:linearized}, the same cannot be said about second-order perturbations. Around a Minkowski background these obey \cite{Arrechea:2020evj,Arrechea:2020gjw}
\begin{equation}
\begin{aligned}
0=&\sbr{\mathrm{GR\,\,terms\,\,of\,\,}\mathcal{O}\cbr{h^2}} + \frac{\alpha}{\d4} \Big[-2\nabla_{\gamma}\nabla_{\alpha}h_{\nu\beta}\nabla^{\gamma}\nabla^{\beta}h_{\mu}{}^{\alpha}
 +2\nabla_{\gamma}\nabla_{\beta}h_{\nu\alpha}\nabla^{\gamma}\nabla^{\beta}h_{\mu}{}^{\alpha}\\& +2\nabla^{\gamma}\nabla^{\beta}h_{\nu}{}^{\alpha}\nabla_{\mu}\nabla_{\alpha}h_{\beta\gamma}
 +2\nabla^{\gamma}\nabla^{\beta}h_{\mu}{}^{\alpha}\nabla_{\nu}\nabla_{\alpha}h_{\beta\gamma}-2\nabla^{\gamma}\nabla^{\beta}h_{\mu}{}^{\alpha}\nabla_{\nu}\nabla_{\beta}h_{\alpha\gamma} 
 \\&-2\nabla^{\gamma}\nabla^{\beta}h_{\nu}{}^{\alpha}\nabla_{\mu}\nabla_{\beta}h_{\alpha\gamma}
 -2\nabla_{\mu}\nabla^{\gamma}h^{\alpha\beta}\nabla_{\nu}\nabla_{\beta}h_{\alpha\gamma}
 +2\nabla_{\mu}\nabla^{\gamma}h^{\alpha\beta}\nabla_{\nu}\nabla_{\gamma}h_{\alpha\beta}\\&+\eta_{\mu\nu}\big(
 2\nabla_{\delta}\nabla_{\beta}h_{\alpha\gamma}\nabla^{\delta}\nabla^{\gamma}h^{\alpha\beta}
 -\nabla_{\delta}\nabla_{\gamma}h_{\alpha\beta}\nabla^{\delta}\nabla^{\gamma}h^{\alpha\beta}-\nabla_{\beta}\nabla_{\alpha}h_{\gamma\delta}\nabla^{\delta}\nabla^{\gamma}h^{\alpha\beta}\big)\Big],
\end{aligned}
\end{equation}
which can be seen to be ill-defined in the four-dimensional limit.

A hint that the original approach outlined by Glavan \& Lin may be an incomplete description of a more complicated theory is given by an analysis of tree-level scattering amplitudes. These reveal that, albeit being different to those of GR, the ones obtained from the four-dimensional limit of the Glavan \& Lin approach are not new. Instead, they all come from certain scalar-tensor theories, indicating the likely presence of a scalar degree of freedom, in addition to the two tensor degrees of freedom in the graviton \cite{Bonifacio:2020vbk}. Moreover, the on-shell action can be observed to contain divergences \cite{Mahapatra:2020rds}, and the field equations cannot be variationally completed in $D=4$, as the Lagrangian diverges \cite{Hohmann:2020cor}.

Given the concerns discussed above, alternative regularizations have been sought for a well-defined version of the Einstein-Gauss-Bonnet theory in four-dimensions. These have resulted in novel scalar-tensor theories either via a conformal regularization \cite{Fernandes:2020nbq,Hennigar:2020lsl} (first applied in Ref. \cite{Mann:1992ar} in two-dimensions), or via a regularized Kaluza-Klein reduction \cite{Lu:2020iav,Kobayashi:2020wqy}. Later, it was shown that these approaches result in scalar-tensor theories included in the subset of Horndeski theories whose scalar-field has improved conformal properties \cite{Fernandes:2021dsb}. Yet another regularization method focuses on temporal diffeomorphism breaking, instead of the inclusion of a scalar degree of freedom \cite{Aoki:2020lig}. We will review each of these approaches in the following sections, but note that in all cases Lovelock's theorem is respected, contrary to the aim of Glavan \& Lin \footnote{Note that ``regularization'', as used here, refers to the process of producing regular field equations in the limit $D\rightarrow 4$, and not to the process of removing divergences in boundary terms.}.

\subsection{Counter-term regularization}
\label{sec:counter-termreg}

The regularization procedure we wish to employ in this section, to create a well-defined version of 4D EGB, was first applied in two space-time dimensions in Ref. \cite{Mann:1992ar} \footnote{Note that in two-dimensions the Ricci scalar has a topological character, much like the Gauss-Bonnet term in four-dimensions, so the problem has a similar character.}. We intend to first go through this procedure in two-dimensions as an instructional demonstration of the methodology that will be used to regularize 4D EGB, which will then follow.

\subsubsection{Regularization in 2D.}
We start by considering the following action in $D$ dimensions:
\begin{equation}
S=\frac{\alpha}{\d2} \int d^Dx \sqrt{-g} R + S_M \,,
\label{eq:2daction}
\end{equation}
where a re-scaling of the coupling constant has been introduced in order to try and cancel the vanishing contribution that $R$ gives to the field equations in $D=2$. Now, if one were to insist on going down the Glavan \& Lin route (i.e. varying the action, obtaining the equations of motion, and then taking the $2D$ limit), one would stumble upon similar problems as in the previously discussed four-dimensional case. This is because the Einstein-tensor is identically zero in two dimensions, and so the limit
\begin{equation}
\lim_{D\to 2} \left( \alpha \frac{G_{\mu \nu}}{\d2}-T_{\mu \nu}\right)\,,
\end{equation}
is not well-defined. Note, however, that the trace of the field equations
\begin{equation}
\lim_{D\to 2} g^{\mu \nu} \cbr{\alpha \frac{G_{\mu \nu}}{\d2}-T_{\mu \nu}} = -\frac{1}{2} \cbr{\alpha R - 2 T}=0,
\label{eq:trace2d}
\end{equation}
is well-defined, just as in the case of 4D EGB. One can attempt to solve this indeterminacy of the field equations by adding a counter-term to the action, in order to cancel the resulting ill-defined terms. This can be done in the present case by adding to the action \eqref{eq:2daction} the term \cite{Mann:1992ar}
\begin{equation}
-\frac{\alpha}{\d2} \int d^Dx \sqrt{-\tilde{g}} \tilde{R} \, ,
\end{equation}
where the tilde denotes a quantity constructed from the conformal geometry defined by
\begin{equation}
\tilde{g}_{\mu \nu} = e^{2\phi} g_{\mu \nu} \, .
\label{eq:confgeo}
\end{equation}
One may note that in $D$ dimensions the square root of the determinant of the metric is related to its conformal counterpart by $\sqrt{-\Tilde{g}}=e^{D \phi}\sqrt{-{g}}$, and that the Ricci scalar of the conformal metric can be specified as \cite{Carneiro:2004rt,Dabrowski:2008kx}
\begin{equation}
\sqrt{-\Tilde{g}} \Tilde{R} = \sqrt{-g} e^{(D-2)\phi} \Big[ R - 2(D-1)\dal - (D-1)(D-2)\dpp \Big] \, .
\label{eq:confR}
\end{equation}
Substituting this all into the action produces the result
\begin{equation}
\begin{aligned}
S&=\frac{\alpha}{\d2} \int d^Dx\sbr{\sqrt{-g} R- \sqrt{-\tilde{g}} \tilde{R}}+S_M \\&= \frac{\alpha}{\d2} \int d^Dx \sqrt{-g} \big[2\d1 \dal +\d1 \d2\dpp -\d2 \phi R \\&\hspace{8cm}+ 2\d2 \d1 \phi \dal\big]+S_M \,,
\end{aligned}
\end{equation}
where we have expanded the exponential around $D=2$ and discarded terms of order $\mathcal{O}(\d2^2)$ or higher. After performing an integration by parts, and discarding boundary terms, we find that the factors of $(D-2)$ cancel, allowing us to take the two-dimensional limit
\begin{equation}
\begin{aligned}
S&=-\frac{\alpha}{\cancel{\d2}}\int d^Dx \sqrt{-g} \cancel{\d2} \cbr{\phi R + \d1 \dpp} +S_M\\& \to -\alpha\int d^2x \sqrt{-g} \cbr{\phi R + \dpp}+S_M \, .
\label{eq:2Daction}
\end{aligned}
\end{equation}
This action has field equations $\tilde{R} = 0$, which are equivalent to
\begin{equation}
R - 2\dal= 0 \, ,
\label{eq:2Dscalareq}
\end{equation}
and
\begin{equation}
\nabla_\mu \phi \nabla_\nu \phi - \nabla_\mu \nabla_\nu \phi + g_{\mu \nu} \left( \dal - \frac{1}{2} \dpp \right) = \frac{1}{\alpha} T_{\mu \nu} \, ,
\label{eq:2Dfeqs}
\end{equation}
where the stress-energy tensor obeys the conservation equation $\nabla^\mu T_{\mu \nu} = 0$. Of particularly interest in this case is that a suitable combination of the scalar field equation \eqref{eq:2Dscalareq} and the trace of the field equations \eqref{eq:2Dfeqs},
\begin{equation}
\dal = \frac{1}{\alpha} T \, ,
\end{equation}
completely decouples from the scalar field resulting in
\begin{equation}
\alpha R=2 T \,,
\label{eq:2Dtrace}
\end{equation}
where $T=T^{\mu}_{\phantom{\mu} \mu}$, thus having the same trace equation as Eq. \eqref{eq:trace2d}. Note that in two dimensions there is only a single degree of freedom in the geometry, which means that Eq. \eqref{eq:2Dtrace} contains all of the information about the theory.

\subsubsection{Regularization in 4D.} 
In this section we apply the ideas from the discussion above to the four-dimensional case with a Gauss-Bonnet term. We start by considering the Einstein-Gauss-Bonnet action in $D$ dimensions
\begin{equation}
S=\int d^D x \sqrt{-g} \cbr{R+\frac{\alpha}{\d4} \mathcal{G}},
\end{equation}
to which we add the counter-term
\begin{equation}
-\frac{\alpha}{\d4}\int d^Dx \sqrt{-\tilde{g}} \tilde{\mathcal{G}}\, ,
\end{equation}
where again the tilde denotes quantities constructed from a conformal geometry as in Eq. \eqref{eq:confgeo}. We can write the Gauss-Bonnet term of the conformal metric in terms of the original one as \cite{Carneiro:2004rt,Dabrowski:2008kx}
\begin{equation}
    \begin{aligned}
\sqrt{-\Tilde g} \Tilde{\mathcal{G}} =&\sqrt{-g} e^{(D-4) \phi}\left[\mathcal{G}-8(D-3) R^{\mu \nu}\left(\nabla_{\mu} \phi \nabla_{\nu} \phi-\nabla_{\mu} \nabla_{\nu} \phi\right) \right. -2(D-3)(D-4) R\dpp \\&+4(D-2)(D-3)^{2}\dal\dpp -4(D-2)(D-3)\left(\nabla_{\mu} \nabla_{\nu} \phi\right)\left(\nabla^{\mu} \nabla^{\nu} \phi\right)\\
&+4(D-2)(D-3)\left(\dal\right)^2 +8(D-2)(D-3)\left(\nabla_{\mu} \phi \nabla_{\nu} \phi\right)\left(\nabla^{\mu} \nabla^{\nu} \phi\right)\\
&-4(D-3) R\dal \left.+(D-1)(D-2)(D-3)(D-4)(\nabla \phi)^4\right].
\end{aligned}
\label{eq:confGauss-Bonnet}
\nonumber
\end{equation}
Expanding the exponential around $D=4$, and neglecting terms of order $(D-4)^2$ or higher, we then obtain
\begin{equation}
\begin{aligned}
    \sqrt{-\Tilde g} \Tilde{\mathcal{G}} = &\sqrt{-g} \Big( \mathcal{G}-4(D-3)R\dal + 4(D-3)^2(D-2)\dal \dpp + 4(D-3)(D-2)(\dal)^2\\ 
    &- 8(D-3)R^{\mu \nu}(\nabla_\mu \phi \nabla_\nu \phi - \nabla_\mu \nabla_\nu \phi)+8(D-3)(D-2)\nabla_\mu \phi \nabla_\nu \phi \nabla^\mu \nabla^\nu \phi \\
    &- 4(D-3)(D-2)(\nabla_\mu \nabla_\nu \phi)(\nabla^\mu \nabla^\nu \phi)+\left(D-4\right) \Big[\phi \mathcal{G} -2(D-3)R\dpp \\&+(D-3)(D-2)(D-1)(\nabla \phi)^4 -4(D-3)\phi R \dal + 4(D-3)^2(D-2)\phi \dal \dpp\\&+4(D-3)(D-2)\phi (\dal)^2 - 8(D-3)\phi R^{\mu \nu}(\nabla_{\mu} \phi \nabla_{\nu} \phi-\nabla_{\mu} \nabla_{\nu} \phi) \\&+ 8(D-3)(D-2)\phi (\nabla_{\mu} \phi \nabla_{\nu} \phi)(\nabla^{\mu} \nabla^{\nu} \phi)- 4(D-3)(D-2)\phi (\nabla_{\mu} \nabla_{\nu} \phi)(\nabla^{\mu} \nabla^{\nu} \phi) \Big] \Big) \, .
\end{aligned} \nonumber
\end{equation}
Integrating by parts, and making use of the identity
$ \nabla_\mu \left[\Box \phi \nabla^\mu \phi - \frac{1}{2} \nabla^\mu \dpp \right] = \left( \Box \phi \right)^2  - \left(\nabla_\mu \nabla_\nu \phi\right)^2 - R^{\mu \nu} \nabla_\mu \phi \nabla_\nu \phi, $
and the Bianchi identities, we can find that the action reads
\begin{equation}
\begin{aligned}
    S=\int_\mathcal{M} d^D x \sqrt{-g} \Big[ R +& \frac{\alpha}{\cancel{\d4}} \cancel{(D-4)} \Big(4(D-3)G^{\mu \nu}\nabla_\mu \phi \nabla_\nu \phi - \phi \mathcal{G}\\&  - 4(D-5)(D-3)\Box \phi \dpp - (D-5)(D-3)(D-2) (\nabla \phi)^4 \Big) \Big] + S_m \, .
\end{aligned} \nonumber
\end{equation}
On taking the four-dimensional limit, this becomes
\begin{equation}
    S=\int_{\mathcal{M}} d^4 x \sqrt{-g} \Big[ R + \alpha \Big(4G^{\mu \nu}\nabla_\mu \phi \nabla_\nu \phi - \phi \mathcal{G} + 4\Box \phi (\nabla \phi)^2 + 2(\nabla \phi)^4\Big) \Big] + S_m \, ,
    \label{eq:4Daction1}
\end{equation}
which can be seen to be a four-dimensional action free of divergences. This action belongs to the Horndeski class of theories \cite{Horndeski,Kobayashi:2019hrl}, with functions $G_2=8 \alpha X^2$, $G_3=8 \alpha X$, $G_4=1+4 \alpha X$ and $G_5 = 4 \alpha \ln X$ (where $X=-\frac{1}{2} \nabla_{\mu} \phi \nabla^{\mu} \phi$). It can also be noted that the action has a shift symmetry in the scalar field, \textit{i.e.}, invariance under the set of transformations $\phi \to \phi + C$, where $C$ is an arbitrary constant.

Given the four-dimensional action \eqref{eq:4Daction1}, the variational principle can be applied to get the field equations
\begin{equation}
G_{\mu \nu} + \alpha \mathcal{H}_{\mu \nu} = T_{\mu \nu},
\label{feqs}
\end{equation}
where
\begin{equation}
\begin{aligned}
\mathcal{H}_{\mu\nu} =& 2G_{\mu \nu} \dpp+4P_{\mu \alpha \nu \beta}\left(\nabla^\alpha \phi \nabla^\beta \phi - \nabla^\beta \nabla^\alpha \phi \right)+4\left(\nabla_\mu \phi \nabla_\nu \phi - \nabla_\nu \nabla_\mu \phi\right) \dal \\
&+4\left(\nabla_\a \phi \nabla_\mu \phi - \nabla_\alpha \nabla_\mu \phi\right) \left(\nabla^\a \phi \nabla_\nu \phi - \nabla^\a \nabla_\nu \phi\right)\\
& +g_{\mu \nu} \Big(2\left(\dal\right)^2 - \left( \nabla \phi\right)^4 + 2\nabla_\b \nabla_\a\phi\left(2\nabla^\a \phi \nabla^\b \phi - \nabla^\b \nabla^\a \phi \right) \Big),
\end{aligned}
\label{eq:TensorH}
\end{equation}
with
$$
P_{\alpha \beta \mu \nu} \equiv *R*_{\alpha \beta \mu \nu} = -R_{\alpha \beta \mu \nu}-g_{\alpha \nu} R_{\beta \mu}+g_{\alpha \mu} R_{\beta \nu}-g_{\beta \mu} R_{\alpha \nu}+g_{\beta \nu} R_{\alpha \mu}-\frac{1}{2}\left(g_{\alpha \mu} g_{\beta \nu}+g_{\alpha \nu} g_{\beta \mu}\right) R,
$$
being the double dual of the Riemann tensor. The corresponding scalar field equation is $\tilde{\mathcal{G}}=0$, which is equivalent to
\begin{equation}
R^{\mu \nu} \nabla_{\mu} \phi \nabla_{\nu} \phi - G^{\mu \nu}\nabla_\mu \nabla_\nu \phi - \dal \dpp +(\nabla_\mu \nabla_\nu \phi)^2 - (\dal)^2 - 2\nabla_\mu \phi \nabla_\nu \phi \nabla^\mu \nabla^\nu \phi = \frac{1}{8}\mathcal{G}.
\label{eq:4Dsfeq}
\end{equation}
Interestingly, a suitable combination of the scalar field equation and the trace of the field equations results in the purely geometric condition,
\begin{equation}
R+\frac{\alpha}{2} \mathcal{G}= -T,
\label{eq:traceeq}
\end{equation}
which is exactly the same trace equation found in the paper by Glavan \& Lin \cite{Glavan:2019inb}. The theory also shares solutions with the original formulation of Ref. \cite{Glavan:2019inb} as we will discuss in later sections.

\subsubsection{Regularization with the dimensional derivative.}
Let us again consider the two-dimensional case, but where we employ the following regularization scheme for the Ricci scalar
\begin{equation}
    S=\alpha \lim_{D\to 2} \frac{\int d^Dx \sqrt{-\tilde{g}}\tilde{R} - \int d^2x \sqrt{-\tilde{g}}\tilde{R}}{\d2} \, .
\end{equation}
This looks very similar to the previously presented regularization, but differs as we add a counter-term whose numerator is already evaluated in two dimensions, where all quantities are tilded. A more careful reading reveals that the expression above is nothing but a dimensional derivative:
\begin{equation}
    S=\alpha \lim_{D \rightarrow 2}\int d^Dx \frac{d}{dD} \left( \sqrt{-\tilde{g}}\tilde{R}\right) \,,
    \label{eq:dimderiv1}
\end{equation}
which one can immediately see to be free of divergences as there is no divergent factor. The dimensional derivative here plays the role of canceling the $\d2$ factors appearing in the equations of motion, much like the divergent factors introduced by Glavan \& Lin \cite{Glavan:2019inb}. That is, 
\begin{equation}
 \lim_{D\to N}   \frac{d}{dD} \cbr{D-N} = 1 \qquad {\rm instead \; of} \qquad \lim_{D\to N} \frac{1}{\cbr{D-N}} \cbr{D-N} = 1.
\end{equation}
To compute the action \eqref{eq:dimderiv1} we proceed by going from the tilded frame to the non-tilded frame
\begin{equation}
\begin{aligned}
    S=&\alpha \lim_{D \rightarrow 2} \int d^Dx \frac{d}{dD} \cbr{\sqrt{-g} e^{(D-2)\phi} \Big[ R - 2(D-1)\dal - (D-1)(D-2)\dpp \Big]} ,
    \\=&\alpha \lim_{D\to 2}\int d^Dx \sqrt{-g} \bigg(e^{(D-2)\phi}\phi \sbr{R - 2(D-1)\dal - (D-1)(D-2)\dpp} 
    \\& \hspace{5cm}- e^{(D-2)\phi} \cbr{2\dal + \cbr{\d1+\d2}\dpp}\bigg) \,,
\end{aligned}
\end{equation}
where we assumed that non-tilded quantities do not possess a dimensional dependence. Evaluating the two-dimensional limit, we observe no divergences and obtain
\begin{equation}
    S=\alpha \int d^2x \sqrt{-g} \cbr{\phi R + \dpp},
\end{equation}
which is the same exact action as we obtained in Eq. \eqref{eq:2Daction} (apart from an overall sign that does not influence the equations of motion). The same dimensional derivative procedure can be applied in 4D to the Gauss-Bonnet invariant, resulting in the same action as the counter-term regularization of Eq. \eqref{eq:4Daction1}. 

\subsection{Regularized Kaluza-Klein reduction}
\label{sec:KKreg}
In this section we review the regularization method employed in Refs. \cite{Lu:2020iav,Kobayashi:2020wqy}, which consists of performing a Kaluza-Klein compactification of $D-$dimensional Einstein-Gauss-Bonnet gravity on a maximally symmetric space-time of $\d4$ dimensions. Here the coupling factor $\alpha$ is taken to have the same singular scaling, and we keep only the breathing mode characterizing the size of the internal space. 

We start the Kaluza-Klein regularization process by parametrizing the $D-$dimensional metric as
\begin{equation}
    ds^2_D=ds^2_4+e^{-2\phi}d\Sigma^2_{D-4,\,\lambda}\,,
\end{equation}
where the scalar field $\phi$ depends only on the $4$-dimensional coordinates, $ds_4^2$ is the $4$-dimensional line-element and $d\Sigma^2_{D-4,\,\lambda}$ is the line-element of an internal maximally symmetric space of $(D-4)$ dimensions whose curvature tensor is given by
\begin{equation}
    R_{abcd}= \lambda(g_{ac}g_{bd}-g_{ad}g_{bc}),
\end{equation}
with $\lambda$ a constant representing the curvature of the internal space.
Under these assumptions, the Einstein-Gauss-Bonnet action \eqref{eq:actionEinstein-Gauss-Bonnet} reduces to
\begin{equation}
    \begin{aligned}
    S=&\frac{1}{16\pi G}\int d^4x\sqrt{-g}e^{-(D-4)\phi}\Bigg\{R+(D-4)(D-5)\big(\dpp+\lambda e^{2\phi}\big)\\&
+\alpha\Big(\mathcal{G}-2(D-4)(D-5)\left[2G^{\mu\nu}\nabla_\mu\phi \nabla_\nu\phi-\lambda R e^{2\phi}\right]\\&-(D-4)(D-5)(D-6)\bigg[-2\dpp \dal+(D-5)(\nabla \phi)^4\bigg]\\
&+(D-4)(D-5)(D-6)(D-7)\left[2\lambda \dpp e^{2\phi}+\lambda^2e^{4\phi}\right]\Big)\Bigg\}\,.
    \end{aligned}
\end{equation}
As we are interested in taking the limit $D\to 4$ we employ a method similar to counter-term regularization, namely expand the exponential and discard terms of order $\d4^2$. Moreover, we can remove the bare Gauss-Bonnet term by introducing a counter-term and re-scaling the coupling constant as $\alpha \to \alpha/\d4$. In the end, the $D\to 4$ limit leaves
\begin{equation}
\begin{aligned}
S=\int d^4x\sqrt{-g}\Big[&R+\alpha \Big(4G^{\mu \nu}\nabla_\mu \phi \nabla_\nu \phi - \phi \mathcal{G} + 4\Box \phi (\nabla \phi)^2 + 2(\nabla \phi)^4\\&\hspace{3cm}-2\lambda e^{2\phi}\sbr{R+6\dpp +3\lambda e^{2\phi}} \Big) \Big].
\end{aligned}
\end{equation}
This regularized action can be seen to differ from the one obtained via the counter-term regularization by the terms proportional to $\lambda$, with $\lambda=0$ (flat internal space) recovering the counter-term regularized theory precisely. 

If one were to keep the $\lambda$-dependent terms, then the resulting field equations would take the form
\begin{equation}
\begin{aligned}
&G_{\mu \nu} + \alpha \left(\mathcal{H}_{\mu \nu} - 2\lambda e^{2\phi} \mathcal{A}_{\mu \nu} +3\lambda^2 e^{4\phi} g_{\mu \nu} \right)= T_{\mu \nu}\, , 
\end{aligned}
\end{equation}
where
\begin{equation}
\begin{aligned}
\mathcal{A}_{\mu \nu} := G_{\mu \nu} + 2\nabla_\mu \phi \nabla_\nu \phi - 2\nabla_\mu \nabla_\nu \phi +g_{\mu \nu} \left( 2\dal + \dpp \right) \, ,
\end{aligned}
\label{eq:TensorA}
\end{equation}
and the scalar field equation takes the form
\begin{equation}
\begin{aligned}
    &4\lambda \cbr{6\lambda + e^{-2\phi} \cbr{R - 6 \dal -6 \dpp}}+e^{-4\phi} \Big(8R^{\mu \nu} \nabla_{\mu} \phi \nabla_{\nu} \phi -8 G^{\mu \nu}\nabla_\mu \nabla_\nu \phi\\& - 8\dal \dpp +8(\nabla_\mu \nabla_\nu \phi)^2 - 8(\dal)^2 - 16\nabla_\mu \phi \nabla_\nu \phi \nabla^\mu \nabla^\nu \phi - \mathcal{G}\Big) = 0 \, .
\end{aligned}
\label{eq:sfeq2}
\end{equation}
Again, the theory possesses the purely geometrical equation given in \eqref{eq:traceeq}, and shares solutions with the original formulation of Ref. \cite{Glavan:2019inb}.

\subsection{Gravity with a generalized conformal scalar field}
\label{sec:generalizedcsf}

\par The structure of the regularized theories described above is highly non-trivial, comprising a representative of each one of the Horndeski terms, and yet they possess an extremely simple field equation (\ref{eq:traceeq}) that completely decouples from the scalar field, and which therefore allows for simple solutions. One is then left to wonder about the relationship that connects these threads, and why it should be that a special combination of the field equations completely decouples from the scalar field. This problem was addressed in Ref. \cite{Fernandes:2021dsb}.

\par A careful analysis of the scalar field equations \eqref{eq:4Dsfeq} and \eqref{eq:sfeq2} reveals that they are conformally invariant, where we define a conformal transformation as
\begin{equation}
    g_{\mu \nu} \to g_{\mu \nu} e^{2\sigma}, \qquad {\rm and} \qquad \phi \to \phi - \sigma.
    \label{eq:weyltransf}
\end{equation}
Conformal symmetry of the matter fields is well-known to be associated with simplifications of the field equations. If the matter action is conformally invariant, then the trace of its stress-energy tensor vanishes, and the theory possesses a constant Ricci scalar. This is the case with electrovaccum, where the Maxwell action is conformally invariant, and is the property that allows us to find the Reissner-Nordstr\"om black hole (and its rotating generalization, the Kerr-Newman black hole) with relative simplicity, despite the presence of a matter source. A further example is gravity with a conformally coupled scalar field, whose matter action enjoys conformal invariance, and is of the well-known form
\begin{equation}
\int d^4x \sqrt{-g} \left(\frac{R}{6} \Phi^2 +\left( \nabla \Phi\right)^2\right).
\label{eq:confcoupledaction}
\end{equation}
The simplification of the field equations due to the conformal symmetry of the matter terms allowed for the finding of the first counter-example to the no-hair theorems (see e.g. Ref. \cite{Herdeiro:2015waa} for a review), the much-debated Bocharova-Bronnikov-Melnikov-Bekenstein (BBMB) black hole \cite{Bocharova:1970skc,Bekenstein:1975ts,Bekenstein:1974sf}. Gravity with a conformally coupled scalar field and its solutions have been the subject of many studies in recent years (see e.g. Refs. \cite{Martinez:2002ru,Martinez:2005di,Padilla:2013jza,deHaro:2006ymc,Dotti:2007cp,Gunzig:2000yj,Anabalon:2009qt,Oliva:2011np,Cisterna:2021xxq,Caceres:2020myr} and references therein).

\par It may be noted that in the above examples the full field equations are not conformally invariant; only the matter field equations such as the Maxwell equations and the modified Klein-Gordon equation resulting from the action \eqref{eq:confcoupledaction} are conformally invariant in each theory. This suggests that the previously mentioned simplification of the equations of motion might in fact be related to the conformal invariance of the matter field equations, and not of the action. Let us analyze in greater detail the consequences of a conformally invariant matter field equation, using the example of a scalar field, following closely Ref. \cite{Fernandes:2021dsb}.

\par We start by considering the transformation of Eq. \eqref{eq:weyltransf} in its infinitesimal form, such that $\delta_\sigma g_{\mu \nu} = 2 \sigma g_{\mu \nu}$ and $\delta_\sigma \phi = -\sigma$, where $\delta_\sigma$ denotes the change under an infinitesimal conformal transformation. Assuming an action principle that describes a theory that depends solely on the metric $g_{\mu \nu}$ and a scalar field $\phi$, $S[\phi,g]$ (such as those that belong to the Horndeski class), we find that the transformation induces the variation
\begin{equation}
\begin{aligned}
\delta_\sigma S &= \int d^4x \left(\frac{\delta S[\phi,g]}{\delta g_{\mu \nu}}\delta_\sigma g_{\mu \nu} + \frac{\delta S[\phi,g]}{\delta \phi} \delta_\sigma \phi\right) \\&= -\int d^4x \left(-2 g_{\mu \nu}\frac{\delta S[\phi,g]}{\delta g_{\mu \nu}} + \frac{\delta S[\phi,g]}{\delta \phi} \right) \sigma,
\end{aligned}
\label{eq:trace1}
\end{equation}
where the first and second terms in brackets can be identified with the trace and the scalar field equations, respectively. Recall that we are analysing the consequences of a theory possessing a conformally invariant scalar field equation; if this is the case, then $\delta_\sigma S$ should be independent of $\phi$, such that the transformed action contains exactly the same scalar field dependence as the original one, resulting in the same scalar field equation (if one were to vary the transformed action, then the scalar field equation should be the same if it is conformally invariant). Thus, the quantity in brackets inside Eq. \eqref{eq:trace1},
\begin{equation}
-2 g_{\mu \nu}\frac{\delta S[\phi,g]}{\delta g_{\mu \nu}} + \frac{\delta S[\phi,g]}{\delta \phi},
\label{eq:geomcomb}
\end{equation}
should be a purely geometric quantity constructed only out of the metric, $g_{\mu \nu}$, and its derivatives. This purely geometric quantity might in general be different from a constant scalar curvature on-shell, while at the same time providing a simple way to find closed-form solutions of the field equations, as we observed with Eq. \eqref{eq:traceeq}. In what follows, we derive the most general scalar-tensor theory with second-order equations of motion and a conformally invariant scalar field equation. Note that scalar quantities constructed solely from the tilded metric are the \textit{only} conformally invariant scalar quantities that depend only of the metric $g_{\mu \nu}$ and the scalar field $\phi$ \cite{Fernandes:2021dsb}. Therefore, the scalar field equation should be constructed only from tilded curvature scalars such that it is invariant under conformal transformations.

\par After establishing this result, our goal is to derive the action principle describing the theory starting from the conformally invariant scalar field equation. This is done in detail in Ref. \cite{Fernandes:2021dsb} and it can be shown that the most general scalar-tensor theory with second-order equations of motion and a conformally invariant scalar field equation is given by the action
\begin{equation}
\begin{aligned}
S=\int d^{4} x \sqrt{-g}\bigg[&R-2\Lambda -\beta e^{2\phi}\left(R + 6(\nabla \phi)^{2}\right)-2\gamma e^{4\phi} \\&+ \alpha \bigg(4 G^{\mu \nu} \nabla_{\mu} \phi \nabla_{\nu} \phi -\phi \mathcal{G} + 4 \square \phi(\nabla \phi)^{2} + 2(\nabla \phi)^{4}\bigg)\bigg]+S_m,
\end{aligned}
\label{eq:actionconfgeneral2}
\end{equation}
for constants $\beta$, $\gamma$ and $\alpha$. The connection to the regularized 4D EGB theories is manifest by setting $\beta=\gamma=0$, while the one obtained via the Kaluza-Klein regularization is recovered for $\beta=2\lambda \alpha$ and $\gamma=3\lambda^2\alpha$. The theory again belongs to the Horndeski class with
\begin{equation}
\begin{aligned}
&G_2 = -2\Lambda -2 \gamma e^{4\phi} + 12\beta e^{2\phi} X + 8\alpha X^2, \qquad G_3 = 8\alpha X, \\&
G_4 = 1-\beta e^{2\phi}+4\alpha X, \qquad G_5 = 4\alpha \log X,
\end{aligned}
\label{eq:horndeskifuncs}
\end{equation}
where $X=-\frac{1}{2}\dpp$. The field equations are
\begin{equation}
G_{\mu \nu} + \Lambda g_{\mu \nu} + \alpha \mathcal{H}_{\mu \nu} - \beta e^{2\phi} \mathcal{A}_{\mu \nu} + \gamma e^{4\phi}g_{\mu \nu} = T_{\mu \nu},
\label{feqs_genconf}
\end{equation}
where $\mathcal{H}_{\mu \nu}$ and $\mathcal{A}_{\mu \nu}$ are defined in Eqs. \eqref{eq:TensorH} and \eqref{eq:TensorA} respectively, and the scalar field equation resulting from the action \eqref{eq:actionconfgeneral2} is equivalent to the vanishing of the tilded (conformally invariant) quantity
\begin{equation}
\beta \tilde{R}+\frac{\alpha}{2} \tilde{\mathcal{G}} + 4\gamma = 0,
\label{eq:sfeq_conf}
\end{equation}
where the tilded quantities are defined in Eqs. \eqref{eq:confR} and \eqref{eq:confGauss-Bonnet} in terms of $g_{\mu \nu}$ and $\phi$. Unsurprisingly, the purely geometric combination \eqref{eq:geomcomb} recovers the condition
\begin{equation}
R+\frac{\alpha}{2} \mathcal{G}-4\Lambda = -T.
\label{eq:traceeq_complete}
\end{equation}
\par The action of Eq. \eqref{eq:actionconfgeneral2} can be reshaped onto a more familiar form via the field redefinition $\Phi=e^{\phi}$, which gives
\begin{equation}
\begin{aligned}
S=\int d^{4} x &\sqrt{-g}\bigg[R-2\Lambda - 6\beta\left(\frac{R}{6}\Phi^2 + \left(\nabla \Phi\right)^2\right)-2\gamma \Phi^4 \\& + \alpha \bigg(\frac{4G^{\mu \nu}\nabla_{\mu} \Phi \nabla_{\nu} \Phi}{\Phi^2} - \log(\Phi) \mathcal{G} + \frac{4\square \Phi(\nabla \Phi)^{2}}{\Phi^3} - \frac{2(\nabla \Phi)^{4}}{\Phi^4}\bigg)\bigg] + S_m,
\label{eq:actionconfgeneral}
\end{aligned}
\end{equation}
where we note the presence of the usual conformally-coupled scalar field action \eqref{eq:confcoupledaction} with a (conformally-invariant) quartic potential. The theory is invariant under the $\mathbb{Z}_2$ symmetry $\Phi \to -\Phi$.

\subsection{Temporal diffeomorphism breaking regularization}

Finally, let us review the alternative regularization of Ref.~\cite{Aoki:2020lig} in which temporal diffeomorphism symmetry is explicitly broken. We begin by performing the Arnowitt-Deser-Misner (ADM) decomposition, for which the metric becomes
\begin{equation}
    ds^2=-N^2 dt^2+\gamma_{ij}(dx^i+N^i dt)(dx^j+N^j dt)\,,
\end{equation}
where $N$ is the lapse function, $N^i$ is the shift vector and $\gamma_{ij}$ is the spatial metric. If one then computes the Einstein-Gauss-Bonnet Hamiltonian in $D=d+1$ dimensions, this results in
\begin{align}
\Hd_{\rm tot}=\int d^dx (N\Hdcal_0+N^i \mathcal{H}_i + \lambda^0 \pi_0+\lambda^i \pi_i)\,, \label{Htot}
\end{align}
where $\lambda^0$ and $\lambda^i$ are Lagrange multipliers, and ($\pi_0$, $\pi_i$, $\pi^{ij}$) are canonical momenta conjugate to ($N$, $N^i$, $\gamma_{ij}$).

To explain the regularization problem in this context, it suffices to analyse the ``energy" part of the Hamiltonian, $\Hdcal_0$. One begins by splitting it into two terms, the regular part, $\Hdcal_{\rm reg}$ and the Weyl part, $\Hdcal_{\rm Weyl}$:
\begin{equation}
\Hdcal_0=\Hdcal_{\rm reg} +\Hdcal_{\rm Weyl} \,.
\end{equation}
After performing the standard substitution $\hat{\alpha}=\alpha/(d-3)$ one can more clearly see the issue using an expansion in powers of $\alpha$, after which, one finds
\begin{align}
\Hdcal_{\rm reg} :=\frac{\sqrt{\gamma}}{2\kappa^2}
\bigg[&
2\Lambda-\Pi-R
+\alpha \bigg\{  \frac{4}{d-2} \Big( R_{ij}R^{ij}-2R_{ij}\Pi^{ij}-\frac{1}{3}\Pi_{ij}\Pi^{ij} \Big)\nonumber\\
&\quad\quad\quad\quad\quad\quad\quad\quad\quad- \frac{d\big( R^2 -2 R\Pi -\frac{1}{3}\Pi^2 \big)}{(d-2)(d-1)} \bigg\} \bigg]
\, + {\cal O}(\hat\alpha^2)
,  \\
\Hdcal_{\rm Weyl} :=
-\frac{\sqrt{\gamma}}{2\kappa^2} &\frac{\alpha}{d-3}
\Big( C_{ijkl}C^{ijkl}-2 C_{ijkl}\Pi^{T\, ijkl}-\frac{1}{3}\Pi^T_{ijkl}\Pi^{T\, ijkl} \Big)
+ {\cal O}(\hat\alpha^2)
, \label{H_Weyl}
\end{align}
where $R_{ij}$ and $R$ are respectively the Ricci tensor and scalar of the $d$-dimensional spatial sections and $\gamma$ is the determinant of the spatial metric. In addition, one represents the conjugate momenta via the variables
\begin{gather}
\Pi^{ijkl}:=8\kappa^4 \Big( \tilde{\pi}^{i[k}-\frac{1}{d-1}\gamma^{i[k} [\tilde{\pi}]  \Big) 
\Big(\tilde{\pi}^{l]j}-\frac{1}{d-1}\gamma^{l]j } [\tilde{\pi}] \Big),
\\
\Pi_{ij}:=\Pi^k{}_{ikj}\,, \quad \Pi:=\Pi^i{}_i
\,,
\end{gather}
with $\tilde{\pi}^{ij}=\pi^{ij}/\sqrt{\gamma}$ and $[\tilde{\pi}]=\gamma_{ij} \tilde{\pi}^{ij}$. The Weyl pieces $C_{ijkl}$ and $\Pi^T_{ijkl}$ are irreducible components of the $d$-dimensional Riemann tensor and of the tensor $\Pi_{ijkl}$, respectively and are  specified by the traceless conditions $C^k{}_{ikj}=0=\Pi^{T\, k}{}_{ikj}$.

Now the regularization problem becomes clear: while the first term, $\Hdcal_{\rm reg}$, appears to converge when $d\rightarrow3$, the Weyl term, $\Hdcal_{\rm Weyl}$, could diverge, or the limit might not be unique. This formulation makes it clear that this term requires the addition of counter-terms to avoid these potential issues.

We now use a simplified example shown in Ref.~\cite{Aoki:2020lig} to further illuminate the issue. Consider a direct product space whose $(d-3)$--dimension component is flat:
\begin{align}
\gamma_{ij}dx^i dx^j=\gamma_{ab}dx^a dx^b +r^2(x^a)\delta_{AB}dx^A dx^B
\,, \label{direct_s}
\end{align}
where the first term corresponds to the 3-dimensional space and the second term corresponds to the $(d-3)$-dimensional space and has a radius $r$ that depends on the position in 3d space, $x^a$.

For this space, one finds the following for several of the components of terms $\Hdcal_{\rm reg}$ and $\Hdcal_{\rm Weyl}$~\cite{Aoki:2020lig}:
\begin{align}\label{scaling}
&\gamma^{ij} R_{ij}=\gamma^{ab} R_{ab} + {\cal O}(d-3), \nonumber\\
&R^{ij} R_{ij}=R^{ab}  R_{ab}+ {\cal O}(d-3), \nonumber\\
&C_{ijkl} C^{ijkl}
=(d-3)\Big( 4R_{ab}R^{ab}-\frac{3}{2}(\gamma^{ab}R_{ab})^2 + F[r,\partial_a r,...]  \Big)
= {\cal O}(d-3)
\,, 
\end{align}
which leads to
\begin{align}
\lim_{d\rightarrow 3} \int d^d x \sqrt{\gamma} N \bigg( \frac{C_{ijkl}C^{ijkl}}{d-3} \bigg)  = {\rm finite}
\,.
\label{Weyl_limit}
\end{align}
The components of $\Hdcal_{\rm reg}$ only depend on the 3-dimensional quantities in the limit $d\rightarrow3$ and are therefore regular, as anticipated. However, while the Weyl term remains finite, it has a dependence on $r(x)$ through the function $F$. This radius $r(x)$ can be interpreted as being the additional degree of freedom found in the other regularization approaches shown above, and indeed this matches the arguments used in the Kaluza-Klein approach. 

In this regularization approach, one is interested in removing additional degrees of freedom via the introduction of counter-terms. The simplest choice of counter-term is the one that just eliminates the function $F[r,...]$ above, but that would just lead to GR. The choice made in Ref.~\cite{Aoki:2020lig} is simply to eliminate the Weyl terms completely. However, the removal of these terms is not covariant, so the end result is a theory that is only invariant under spacial diffeomorphisms and therefore breaks the full 4D diffeomorphism invariance of GR.

In order to fully determine the theory, a gauge-fixing condition is necessary to choose the constant time hypersurfaces that are preferred when breaking the temporal part of the diffeomorphisms. This gauge fixing condition is added to the Hamiltonian via
\begin{align}
\Hd_{\rm tot}'' &= \Hd_{\rm tot}+ \int d^dx \lambda_{\rm GF} \Gdcal(\gamma_{ij},\pi^{ij}) + \Hd_{\rm ct} 
\nonumber\\
&=\Hd_{\rm reg} + \int d^dx \lambda_{\rm GF} \Gdcal(\gamma_{ij},\pi^{ij}),
\end{align}
where $\lambda_{\rm GF}$ is a Lagrange multiplier. This gauge choice is part of the definition of the theory and one could therefore define many such theories with different gauges. The choice made in Ref.~\cite{Aoki:2020lig} is given by $\limitGcal=\sqrt{\gamma} D^2 [\tilde{\pi}]$, where $D_i$ is the spatial covariant derivative. With this, the following Lagrangian is obtained:
\begin{align}
\mathcal{L}^{\rm 4D}_{\rm Einstein-Gauss-Bonnet}&=\frac{1}{2\kappa^2} ( -2\Lambda +\mathcal{K}_{ij}\mathcal{K}^{ij}-\mathcal{K}^i_{\,i}\mathcal{K}^j_{\,j}+R+\hat\alpha R^2_{\rm 4DGauss-Bonnet} )\,, \nonumber\\
R^2_{\rm 4DGauss-Bonnet}&=-\frac{4}{3}\left(8R_{ij}R^{ij}-4R_{ij}\mathcal{M}^{ij}-\mathcal{M}_{ij}\mathcal{M}^{ij}\right) +\frac{1}{2}\left( 8R^2 -4 R\mathcal{M} -\mathcal{M}^2 \right)\,, 
\label{action}
\end{align} 
where 
\begin{gather}
\mathcal{K}_{ij}=\frac{1}{2N}( \dot{\gamma}_{ij}-2D_{(i}N_{j)}-\gamma_{ij}D^2 \lambda_{\rm GF} )\,,\\
\mathcal{M}_{ij}:=R_{ij}+\mathcal{K}^k_{\, k} \mathcal{K}_{ij}-\mathcal{K}_{ik}\mathcal{K}^k_{\, j}\,,\quad \mathcal{M}:=\mathcal{M}^i_{\, i}.
\end{gather}
This theory propagates the same number of degrees of freedom as GR, but breaks time diffeomorphisms and therefore is still in agreement with Lovelock's theorem. The phenomenology of the theory has some differences with respect to the original theory, particularly regarding the propagation of gravitational waves, as will be discussed below \cite{Aoki:2020iwm,Aoki:2020ila}. Nonetheless, the black hole and FLRW solutions of the original theory are present in this framework \cite{Aoki:2020lig}. 


\section{Black Holes} \label{sec:bh}

\par The black holes of the original theory of Ref. \cite{Glavan:2019inb} were briefly discussed in Section \ref{sec:4D EGB}. In this section we will analyse the black hole solutions in more detail, in the original 4D EGB theory as well as in the various regularized alternatives presented above. These solutions are of obvious interest for strong field gravity, and from the recent observations of black holes via gravitational waves and their accretion disks.

Recall that the field equations reveal the solution below, discussed in the introduction,
\begin{equation}
ds^2 = -f(r) dt^2 + \frac{dr^2}{f(r)} + r^2 \cbr{d\theta^2 + \sin^2 \theta d\varphi^2},
\label{eq:sphsymmlineelement}
\end{equation}
\begin{equation}
f(r) = 1+\frac{r^2}{2\alpha} \cbr{1\pm \sqrt{1+\frac{8M\alpha}{r^3}}}.
\label{eq:BHsolution}
\end{equation}
As with maximally symmetric space-times, there are two branches: the Gauss-Bonnet branch with a plus sign, and the GR branch with the minus sign. In the far field limit, $r \to \infty$, the GR branch solution approaches the Schwarzschild solution
\begin{equation}
f(r)  = 1-\frac{2M}{r}+\mathcal{O}\cbr{r^{-2}},
\end{equation}
whereas the Gauss-Bonnet branch is not asymptotically flat
\begin{equation}
    f(r) = \frac{r^2}{\alpha} + 1 + \frac{2 M}{r}+\mathcal{O}\cbr{r^{-2}}.
\end{equation}
Furthermore, in the vanishing-$\alpha$ limit we observe
\begin{equation}
\begin{aligned}
&f(r)  = 1-\frac{2M}{r}+\mathcal{O}\cbr{\alpha}, \quad \mbox{(GR branch)},\\& f(r) = \frac{r^2}{\alpha} + 1 + \frac{2 M}{r}+\mathcal{O}\cbr{\alpha}, \quad \mbox{(Gauss-Bonnet branch).}
\end{aligned}
\end{equation}
Thus, the Gauss-Bonnet branch is typically disregarded as a physically-interesting solution, because it is not asymptotically flat, does not present a well-defined limit as $\alpha$ vanishes and because the mass term has the wrong sign.

\begin{figure}[ht!]
\centering
\includegraphics[width=0.7\textwidth]{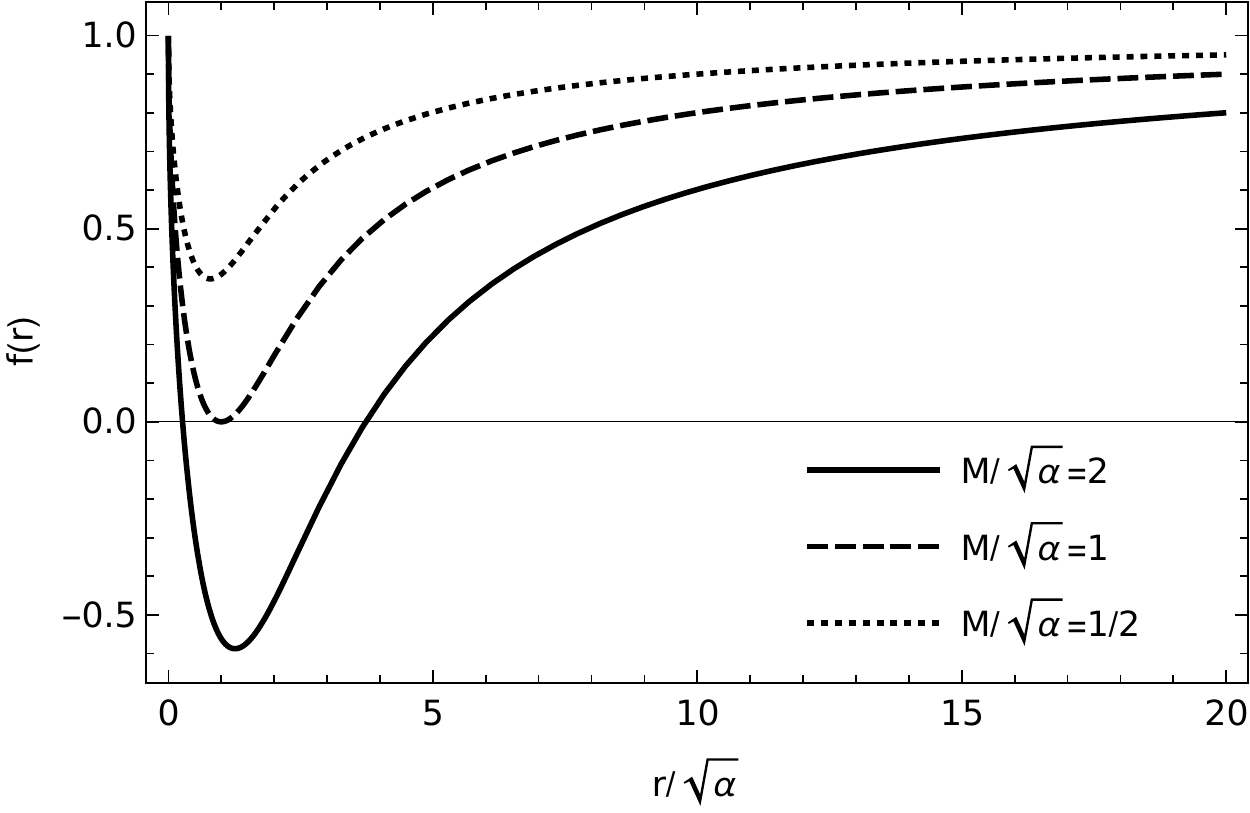}
\caption{The metric function $f(r)$, as a function of the radial coordinate for several values of $M/\sqrt{\alpha}$, and for the minus sign branch.}
\label{fig:BHsolution}
\end{figure}

A closer look at the solution presented in Eq. \eqref{eq:BHsolution} reveals that there are (generically) two horizons located at
\begin{equation}
r_\pm = M \pm \sqrt{M^2-\alpha} \,,
\label{eq:horizons}
\end{equation}
where we note that $r_+r_-=\alpha$ and that the metric components are finite as $r\to 0$:
\begin{equation}
\lim_{r\to 0} f(r) = 1.
\end{equation}
Both these features can be observed in Fig. \ref{fig:BHsolution}.
Even though the metric components are finite, the central singularity located at $r=0$ still exists as the Ricci and Kretschmann scalars behave as
\begin{equation}
R \propto r^{-3/2} \qquad {\rm and} \qquad R_{\mu \nu \alpha \beta}R^{\mu \nu \alpha \beta} \propto r^{-3},
\end{equation}
near $r=0$. Note, however, that the Gauss-Bonnet term has weakened the singularity when compared to the Schwarzschild black hole from GR, where the Kretschmann scalar diverges as $r^{-6}$ near the center. Another singularity exists if we were to consider a negative value of $\alpha$, located at the radius for which the quantity inside the square root in the solution \eqref{eq:BHsolution} vanishes:
\begin{equation}
r^3 = -8 M\alpha. 
\end{equation}
However, the reader should bear in mind that negative values of $\alpha$ are tightly constrained, as we shall discuss in the next sections. On the phenomenology side, the shadow, ISCO and the quasi-normal modes of these black holes have been studied in Refs. \cite{Konoplya:2020bxa,Guo:2020zmf}.

Charged black hole generalizations with a cosmological constant also exist and take the form \cite{Fernandes:2020rpa}
\begin{equation}
f(r) = 1+\frac{r^2}{2\alpha} \cbr{1\pm \sqrt{1+4\alpha \cbr{\frac{2M}{r^3}-\frac{Q^2}{r^4} + \frac{\Lambda}{3}}}},
\label{eq:BHgeneral}
\end{equation}
which is once again very similar to their higher-dimensional generalizations \cite{Wiltshire:1988uq,Wiltshire:1985us,Cai:2001dz} (see Ref. \cite{Garraffo:2008hu} for a review on higher-dimensional Lovelock black holes). In this case, in the abscence of the cosmological constant, the horizons are located at
\begin{equation}
r_\pm = M \pm \sqrt{M^2-Q^2-\alpha}\,.
\end{equation}
It is interesting to note that the metric function in Eq. \eqref{eq:BHgeneral} can be written in dimensionless form by introducing the dimensionless quantities/variables
\begin{equation}
\tilde{r}\equiv \frac{r}{r_+}, \qquad \tilde{\alpha} \equiv \frac{\alpha}{r_+^2}, \qquad \tilde{Q}^2 \equiv \frac{Q^2}{r_+^2} \qquad \tilde{\Lambda} \equiv \Lambda r_+^2,
\end{equation}
such that
\begin{equation}
f(\tilde{r}) = 1+\frac{\tilde{r}^2}{2\tilde{\alpha}}\cbr{1\pm \sqrt{1+4\tilde{\alpha} \cbr{\frac{1+\tilde{\alpha}}{\tilde{r}^3}-\frac{\tilde{Q}^2}{\tilde{r}^4}\cbr{\tilde{r}-1}+\frac{\tilde{\Lambda}}{3}\cbr{1-\frac{1}{\tilde{r}^3}}} }}.
\end{equation}
Let us now consider how the form of black hole solution in the various alternative 4D EGB theories from Section \ref{sec:4D EGB}.

\subsection{Black Hole solutions of the regularized scalar-tensor theories}

For the scalar-tensor regularized theories we will discuss solutions of the field equtions that arise from the action in Eq. \eqref{eq:actionconfgeneral}, as all other theories are subsets for appropriately chosen couplings $\beta$ and $\gamma$ (and $\Lambda$). We supplement the theory with an electromagnetic field for the sake of generality, thus our action is
\begin{equation}
S=\frac{1}{16\pi} S_{\mathcal{G}\Phi} - \frac{1}4 \int d^{4} x \sqrt{-g} F_{\mu \nu} F^{\mu \nu},
\end{equation}
where $S_{\mathcal{G}\Phi}$ is the action defined in Eq. \eqref{eq:actionconfgeneral}. We start by employing a generic static and spherically-symmetric line-element given by Eq. \eqref{eq:sphsymmlineelement}, along with a four-potential
\begin{equation}
    A_\mu dx^\mu = V(r) dt - \frac{Q_m}{4\pi} \cos{\theta} d\varphi,
\end{equation}
where $Q_m$ is interpreted as the magnetic charge of the solution. The Maxwell equations are straightforward to solve and reveal that
\begin{equation}
    V(r) = -\frac{Q_e}{4\pi r} - \Psi_e,
\end{equation}
where $Q_e$ is interpreted as the electric charge of the solution and $\Psi_e$ is the difference in electrostatic potential between the event horizon and infinity. For future convenience we define
\begin{equation}
\mathcal{Q}^2=\frac{Q_e^2+Q_m^2}{4\pi}.
\label{eq:reducedcharge}
\end{equation}

We note that the purely geometrical field equation is independent of $\beta$ and $\gamma$, which thus for the above line-element takes the remarkably simple form \cite{Fernandes:2021dsb}
\begin{equation}
    R+\frac{\alpha}{2}\mathcal{G}-4\Lambda = r^{-2} \sbr{\cbr{1-f}\cbr{r^2+\alpha \cbr{1-f}}}''-4\Lambda = 0
\end{equation}
with the primes denoting radial derivatives. This equation can be integrated to give the general solution
\begin{equation}
    f(r)=1+\frac{r^2}{2\alpha} \sbr{1\pm \sqrt{1+4\alpha \cbr{\frac{2M}{r^3}-\frac{q}{r^4}+\frac{\Lambda}{3}}}},
    \label{eq:metricfunction}
\end{equation}
where $M$ and $q$ are constants. Another particularly interesting field equation results from a suitable combination of the $tt$ and $rr$ field equations, and factorizes into a condition equivalent to
\begin{equation}
\left(\frac{\Phi'}{\Phi^2}\right)'\left(f \Phi' \left(r^2 \Phi\right)' + (f-1)\Phi^2-\frac{\beta}{2\alpha} r^2 \Phi^4\right)=0\,.
\end{equation}
This reveals the possible scalar field profiles to be
\begin{equation}
\begin{aligned}
&\mbox{(1)} \qquad \Phi = \frac{c_1}{r+c_2} \, ,\\&
\mbox{(2)} \qquad \Phi = \frac{\sqrt{-2\alpha/\beta}\, \mbox{sech} \left(c_3 \pm \int^r \frac{dr}{r \sqrt{f}}\right)}{r}\, ,\\&
\mbox{(3)} \qquad  \Phi=c_4\, ,\\&
\mbox{(4)} \qquad \Phi=\frac{\exp\left(c_5\pm \int^r \frac{dr}{r \sqrt{f}} \right)}{r} \quad \mbox{if} \quad \beta=0 \,,
\label{eq:scalarfieldprofile}
\end{aligned}
\end{equation}
where the $c_i$s here are all constants. Let us now consider the theories from Section \ref{sec:4D EGB} as seperate cases.

\textbf{Case of section \ref{sec:counter-termreg}: $\beta=\gamma=0$}\\
Here $\beta=0$, and so we employ the fourth scalar field profile in Eq. \eqref{eq:scalarfieldprofile}. The remaining field equations then reveal that a solution exists for $q=\mathcal{Q}^2$, while $c_5$ in the scalar field is unconstrained, thus being a free parameter (this is due to the shift-symmetry of $\phi$). Ref. \cite{Fernandes:2021ysi}, in particular, studied the asymptotically flat spherically symmetric solutions of this theory in detail, where a uniqueness theorem was obtained for the aforementioned black hole space-time. 

\textbf{Case of section \ref{sec:KKreg}: $\beta=2\lambda \alpha$ and $\gamma=3\lambda^2\alpha$}\\
For the theory resultant from the Kaluza-Klein regularization scheme, we have a solution again given by the metric function \eqref{eq:metricfunction} with $q=\mathcal{Q}^2$ for the scalar field profile number (2) in Eq. \eqref{eq:scalarfieldprofile} (with $\beta=2\lambda \alpha$) \cite{Lu:2020iav}. No other solutions are known in closed form, and numerical results suggest that other spherically symmetric solutions represent naked singularities and not black hole space-times \cite{Lu:2020iav}.

\textbf{Generic $\beta$ and $\gamma$ case of section \ref{sec:generalizedcsf}}\\
For other values of $\beta$ and $\gamma$, black hole solutions were found in Ref. \cite{Fernandes:2021dsb}. First, using the scalar field profile (1) of Eq. \eqref{eq:scalarfieldprofile}, a black hole solution was found for the metric function \eqref{eq:metricfunction} with
\begin{equation}
    q=\mathcal{Q}^2-2\alpha, \qquad c_1=\sqrt{-2\alpha/\beta}, \qquad c_2 = 0, \qquad \frac{\gamma}{\beta^2} = \frac{1}{4\alpha}.
\end{equation}
Using the scalar field profile (2), a black hole solution similar to the Kaluza-Klein case (but with less restrictive couplings) exists. This occurs for a coupling that obeys $\gamma/\beta^2 = 3/4\alpha$, just as in the Kaluza-Klein case, but no specific value is required for either $\beta$ or $\gamma$ as long as the previously stated ratio is obeyed (note that in the Kaluza-Klein case the values $\gamma$ and $\beta$ are fixed).
\par Finally, a \textit{critical solution} with constant scalar field $\Phi=c_4 = \sqrt{1/\beta}$ exists, as long as $\gamma/\beta^2 = -\Lambda$. In this extreme situation the field equations become an identity, and we are left with solving only the purely geometrical condition, whose general solution is given by the metric function of Eq. \eqref{eq:metricfunction} with unconstrained $q$.

\subsection{Slowly Rotating Solutions}
In this section we briefly discuss the slowly-rotating black hole solutions. Fully rotating solutions to the 4DEGB field equations (in any of its formulations) are not yet known in closed form, and it is likely that numerical methods are necessary. However, slowly-rotating solutions can be constructed, as was done originally in Ref. \cite{Charmousis:2021npl} for the scalar-tensor theory of Section \ref{sec:counter-termreg}. Treating rotation as a perturbation to the static solutions, we follow the Hartle-Thorne formalism \cite{Hartle:1967he,Hartle:1968si}. At first order in the rotation parameter we parametrize the metric as
\begin{equation}
   ds^2 = - f(r) dt^2 + \frac{dr^2}{f(r)} + r^2 \left(d\theta^2 + \sin^2\theta d\varphi^2 - 2 \omega(r) \sin^2\theta dt d\varphi \right),
\end{equation}
where $\omega(r)$ is small and of order of the black hole angular velocity, to be determined by the field equations. Plugging the above ansatz into the field equations and expanding in powers of $\omega(r)$, one observes that only the $t\varphi$-component of the field equations is yet to be satisfied, when provided with the background solution discussed above. The $t\varphi$-equation then becomes equivalent to the differential equation
\begin{equation}
    4 \cbr{r^3+5M\alpha} \omega' + r \cbr{r^3+8M\alpha} \omega'' = 0 \,,
\end{equation}
with physically-interesting solution
\begin{equation}
    \omega(r) = \frac{J}{2M\alpha} \cbr{\sqrt{1+\frac{8M\alpha}{r^3}}-1},
\end{equation}
where $J$ is to be interpreted as the angular momentum of the black hole. Again we observe that at large distances we have
\begin{equation}
    \omega(r) = \frac{2J}{r^3} -\frac{4\alpha J M}{r^6} + \mathcal{O}\cbr{r^{-9}},
\end{equation}
and for small coupling $\alpha$ we have
\begin{equation}
    \omega(r) = \frac{2J}{r^3} -\frac{4\alpha J M}{r^6} + \mathcal{O}\cbr{\alpha^2} \,.
\end{equation}
These solutions are therefore identical to GR at leading order in the appropriate limits. A more detailed phenomenological study of slowly rotating solutions is lacking, and presents an avenue for further research.

\subsection{Thermodynamics}
Let us now briefly review some of the thermodynamical properties of the static 4D EGB black holes. We will consider for starters the general metric function of Eq. \eqref{eq:metricfunction}. The temperature of this black hole can be computed to be \cite{Fernandes:2020rpa}
\begin{equation}
    T_+ = \frac{1}{4\pi} f'(r_+) = \frac{r_+^2 \cbr{1-r_+^2 \Lambda}-\alpha -q}{4\pi r_+ \cbr{r_+^2+2\alpha}}\,.
    \label{eq:BHtemp}
\end{equation}
Interestingly, the temperature vanishes as the event horizon approaches a size
\begin{equation}
    r_+^2 = \frac{1-\sqrt{1-4\cbr{q+\alpha}\Lambda}}{2\Lambda} \,,
\end{equation}
which for vanishing cosmological term and $q$ translates into the minimum horizon size $r_+=\sqrt{\alpha}$. The evaporation of 4DEGB black holes was studied in detail in Ref. \cite{Fernandes:2021ysi}, where the idea that evaporation remnants with size $r_+=\sqrt{\alpha}$, from primordial black holes, might constitute all dark matter was put forward. This scenario was found to be compatible with current cosmological observations provided that $\sqrt{\alpha} \lesssim 10^{-18} \mbox{m}$.

\par The entropy of such black holes can be computed via the first law of thermodynamics, $dM = T dS + \sum_i \mu_i dQ_i$, and gives
\begin{equation}
    S= S_0 + \int \frac{dM}{T_+},
\end{equation}
for a constant $S_0$. Using Eq. \eqref{eq:BHtemp}, one can obtain
\begin{equation}
    S = \frac{A}{4} + 2\pi \alpha \log \cbr{\frac{A}{A_0}},
\end{equation}
where $A_0$ is a constant with units of area. This result is valid for any $q$ and $\Lambda$. Logarithmic corrections to the black hole entropy are of interest as they often appear as the leading-order corrections in quantum gravity \cite{Kaul:2000kf,Sen:2012dw}.

\section{Cosmology}\label{sec:cos}

In order to be viable, any theory of gravity must give rise to a cosmology that is both internally consistent, 
and compatible with the myriad of modern cosmological observations across a 
huge range of energy, distance, and time scales. 
 It is also 
interesting to see what new behaviour novel theories might allow, 
for example at very early times where 
observational constraints are less stringent. In this section, we turn to these considerations 
for the 4D Einstein-Gauss-Bonnet theory.

\subsection{Background cosmology}

We begin with the homogeneous and isotopic Friedmann-Roberston-Walker (FRW) line-element in $D$ dimensions
\begin{equation}
{\rm d}s ^2 = - {\rm d}t^2 + a^2(t) \left[{\rm d}\chi^2 + S_k^2(\chi)  {\rm d}\Omega^2 \right ]\,.
\label{eq:FRW}
\end{equation}
where $a(t)$ is the scale factor, $ {\rm d}\Omega^2$ represents the line element for 
an $D-2$ sphere and 
$S_k$ takes the form $S_k(\chi) = \sin(\chi)$ for a positively curved, $k=1$, universe, 
$S_k(\chi) = \chi$  for a flat $k=0$ universe, and $ S_k(\chi) = \sinh(\chi)$ for a negatively curved $k=-1$ universe. 
As was previously described for the flat case in Section \ref{sec:4D EGB}, considering the original theory 
and following the steps discussed there, the space-time given in  
Eq.~\eqref{eq:FRW} with perfect fluid matter source, $T^{\mu}_\nu = (-\rho, p,p,p,\ldots)$, gives rise to the following Friedmann equation in the limit $D\rightarrow 4$:
\begin{equation}
H^2+\frac{k}{a^2}+\alpha \left(H^2+\frac{k}{a^2}\right)^2 = \frac{8\pi G }{3} \rho \,,
\label{eq:Fried1}
\end{equation}
where $H=\dot a/a$, and the density contains all fluids present, i.e. $\rho=\sum_m{\rho_m}$.
We assume any cosmological constant is included in $\rho$, and note that since the stress-energy tensor is conserved that all the components of $\rho$ are expected to obey the same conservation equations as in GR:
\begin{equation}
\dot \rho_m + 3 H (\rho_m + p_m) = 0\,,
\label{eq:Fried3}
\end{equation}
where $p_m$ represents the pressure of the fluids (i.e. we will not consider interacting fluids here).

The regularised theories of Section \ref{sec:4D EGB}  lead to the same equations as \eqref{eq:Fried1}-\eqref{eq:Fried3} when particular forms for the scalar field solution are taken, and particular parameter choices are made. We 
will return to more general solutions, and possible restrictions on solutions in the 
context of regularised theories below, and for now focus on the behaviours prescribed by Eqs.~\eqref{eq:Fried1}-\eqref{eq:Fried3}. 

Considering the Friedmann equation~\eqref{eq:Fried1}, we find that 
\begin{equation}
H^2+\frac{k}{a^2} = \frac{-1 \pm \sqrt{1 + \frac{32 \pi G \alpha \rho}{3}}}{2 \alpha}\,.
\label{eq:Fried4}
\end{equation}
From this equation we see that 
the negative branch
does not lead to a consistent cosmology, but that 
selecting the positive branch leads us to an 
equation that agrees with the standard Friedmann equation 
as $\rho$ tends to zero. 

For negative $\alpha$, $H^2$ becomes complex when $\rho >  3/(32 \pi G |\alpha| )$, and 
hence our universe could not have existed at sufficiently high energies in our past for this case. This 
is problematic if this value of $\rho$ is at or below the inflationary energy scale, placing a 
strong constraint on negative values of $\alpha$ \cite{Clifton:2020xhc} as will be discussed below.
For positive $\alpha$, there is no restriction on the energy scale at which Eq. (\ref{eq:Fried4}) is valid, 
but one may note that the dynamics  can be significantly altered at high energies, 
and hence early times, becoming closer and closer to standard cosmology as $\rho$ decays. 
In this case when $\rho \gg 1/(\alpha G) $ one finds $H^2+\frac{k}{a^2} \propto \sqrt{\rho}$, 
which can have interesting consequences particularly in the positively curved case. 

\begin{figure}[ht!]
\centering
\includegraphics[width=0.7\textwidth]{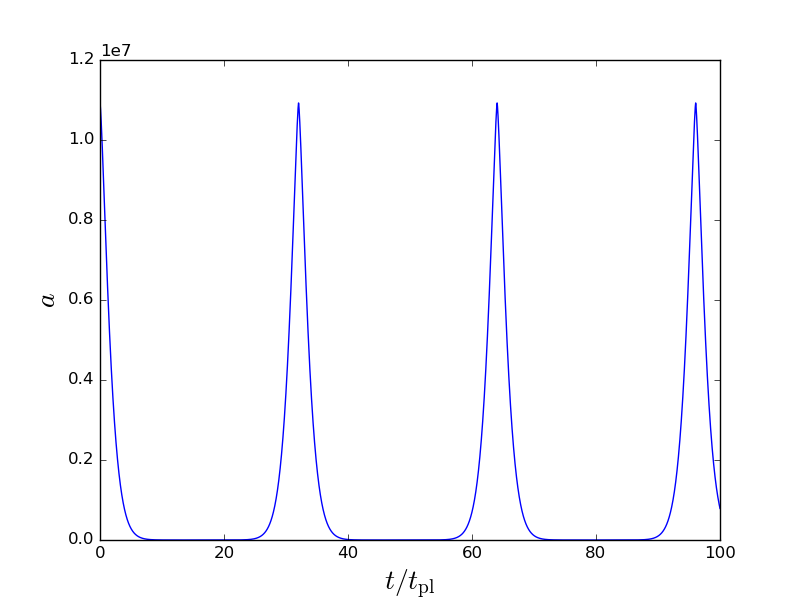}
\caption{The results of numerically solving the equation of motion for $\ddot a$ (which follows from differentiating Eq. \eqref{eq:Fried1}) together with Eq. \eqref{eq:Fried4} for a dust cosmology with $w=1$. We see that the universe undergoes repeating cycles. In this example $\alpha = 100 l_{\rm pl}^2$. The maximum value of $a$ at which the simulation begins is found by fixing the initial density (which can be arbitrarily small), and finding the value of $a$ which 
sets $H^2$ using Eq.~\eqref{eq:Fried4} to be zero.}
\label{fig:cycles}
\end{figure}

Defining the equation of state $w$ through the 
equation $\rho =w p$, the conservation equation \eqref{eq:Fried3} implies $\rho \propto a^{-3(1+ w)}$. This in turn means that for large $\rho$ the right-hand side Eq.~\eqref{eq:Fried4} scales as $a^{-3(1+ w)/2}$. 
For the positively curved case, this tells us that a collapsing universe will undergo a bounce  if it is 
sourced by fluids with a combined equation of state that satisfies the condition $w<1/3$ (recall that $w=1/3$ 
is the equation of state for radiation). This follows because when this condition is met, 
the right-hand side of Eq.~\eqref{eq:Fried4} grows 
more slowly than the curvature term as $a \to 0$. Taking the curvature term to the right-hand side of Eq.~\eqref{eq:Fried4} we can see that it is negative for $k=1$, and if it grows faster than the other,
positive, term, there will come a value of $a$ at which $H^2$ goes to zero. Before this point, in a 
collapsing universe, $H$ is negative, and at this point it passes through zero and 
becomes positive and $H^2$ starts to grow again. Moreover,  once the condition
$\rho \ll 1/G \alpha$ is reached in the expansion phase, the term of the 
right-hand side Eq.~\eqref{eq:Fried4} starts to decay more rapidly than the curvature term, and once again this will lead 
to $H$ passing through zero, and the universe re-collapsing. The result is a cyclic universe. An example is shown in Fig.~\ref{fig:cycles} for the case of a dust cosmology.

Given the appeal of bouncing universes, and the historical interest in them, it is 
intriguing that  4D Einstein-Gauss-Bonnet permits cyclic cosmologies. 
The exotic behaviour described is, however, unlikely to have physical consequences given that the 
phenomena is only apparent when $\rho \alpha G \gg 1$. As we will see below, for the values of 
$\alpha$ allowed by observational constraints, this condition would imply a value of $\rho$ that corresponds to a time 
well into the radiation dominated era, when a bounce cannot occur.

More generally, however, we note that the modified Friedmann equation changes the relationship between 
energy density and the scale factor, changing the universe's expansion history.  
This change becomes significant as $\rho$ approaches $1/(\alpha G)$ at 
 early times, affecting for example the  relation between 
time and temperature. This has consequences for the confrontation of the theory with 
observation, as we will see below.

\subsection{Perturbed FRW cosmology}

In order to study the origin and evolution of structure in the universe, as well as the propagation of 
gravitational waves, it is necessary to introduce perturbations to the line-element \eqref{eq:FRW}, and to the stress-energy tensor. 
Specialising to the flat case and using Poisson gauge~\cite{Malik:2008im}, we have
\begin{equation}
{\rm d}s ^2 = -  a^2(\tau)(1+2\varphi){\rm d}\tau^2  + a^2(\tau) (1-2\psi \delta_{ij}+2 \partial_{(i} F_{j)} + \gamma_{ij}) {\rm d}x_j^2 \,.
\label{eq:FRWpert}
\end{equation}
and 
\begin{equation}
\delta T^0_0 = -\delta \rho \,, ~~~\delta T^0_i = (1+ w) \rho (\partial_i v+v_i) \,, ~~~\delta T^i_j = \delta^i_j \delta P +\pi^i_{j} \,,
\end{equation}
where $\tau$ is conformal time, $\varphi$,  $\psi$, and  $v$, $ \delta \rho$ and $\delta P$ are perturbative scalar quantities, $F_i$ and $v_i$ are transverse vector quantities and $\gamma_{ij}$ is taken to be transverse and 
trace-free, and represents gravitational waves. We have assumed 
a general fluid with anisotropic stress, $\pi$, whose scalar-vector-tensor decomposition is, at first order,
\begin{equation}
\pi_{ij}=(\partial_i\partial_j-\tfrac13\delta_{ij}\partial^2)\Pi+\partial_{(i}\Pi_{j)}+\Pi_{ij} \,.
\end{equation}
This 
line element leads to evolution 
equations for the perturbations which are modified from those of general relativity. 

The propagation equation for gravitational waves \cite{Glavan:2019inb,Feng:2020duo} is given by 
\begin{equation}
\ddot{\gamma}_{ij}+\cbr{3+\frac{4\alpha \dot H}{1+2\alpha H^2}}H \dot{\gamma}_{ij}-c_T^2\frac{\partial^2 \gamma_{ij}}{a^2}=\frac{8\pi G}{{1+2\alpha H^2}}\Pi_{ij}\,,
\label{eq:GWs}
\end{equation}
where $c_T^2 = c^2(1 + \dot{\Gamma}/(H \Gamma)) $ and $\Gamma \equiv 1 + 2 \alpha H^2/c^2 = \sqrt{1 + 32\pi G\rho \alpha/3}$. Neglecting anisotropic stress, this equation also follows from the action
\begin{equation}
\label{eq:tensorf}
S_T= \frac{1}{2}\int {\rm d}^3 x {\rm d} t \,a^3 \Gamma \left ( \dot{\gamma}^2  - c_T^2 \frac{(\partial \gamma )^2}{a^2} \right)\,,
\end{equation}
which will be useful below.

Defining $c_s^2=\delta P/\delta \rho$ and $\delta=\delta \rho/\rho$, the scalar perturbations obey the equations  \cite{Haghani:2020ynl,Feng:2020duo,Wang:2021kuw}
\begin{eqnarray}
\delta' = 3 {\cal H} (w -c_s^2)\delta+(1+w)(3 \psi' - \theta)\,,\nonumber \\
\theta' + \left((1-3 w){\cal H}+\frac{w'}{1+w}\right) \theta + \partial^2 \varphi +\frac{3\partial^2\delta P+2\partial^4\Pi}{3(1+w)\rho}=0\,,\nonumber \\
8 \pi G a^4 \rho \delta + 2 {\cal A} (- \partial^2 \psi + 3 {\cal H}^2 \varphi + 3 {\cal H} \psi') = 0\,, \nonumber \\
{\cal A} \varphi + ({\cal B} - 4 \alpha {\cal H}') \psi =-8\pi G a^4\Pi\,,
\label{eq:scalarPerts}
\end{eqnarray}
where a dash indicates differentiation with respect to conformal time, $\theta =\grad^2 v$, ${\cal A} = 2\alpha {\cal H}^2 +a^2$ and ${\cal B} = 2\alpha {\cal H}^2 -a^2$. The first two equations 
follow from the conservation of the stress-energy tensor and are identical to those of GR. The 
latter two follow from the $00$- and $ij$-components of the field equations, respectively. 
For completeness, we include also the equations for vector modes. They are given by
\begin{eqnarray}
{\cal A}\partial^2F_i'+16\pi G(1+w)a^4\rho v_i=0\,,\nonumber\\
v_i' + \left((1-3 w){\cal H}+\frac{w'}{1+w}\right) v_i +\frac{\partial^2\Pi_i}{2(1+w)\rho}=0\,.
\label{eq:vectorPerts}
\end{eqnarray}

As was the case for the background equations, 
the significance of 
deviations of Eqs.~\eqref{eq:GWs}-\eqref{eq:vectorPerts} from those of GR 
are determined by the size of $\rho G \alpha$ (or equivalently $\alpha H^2$) compared to unity, and hence 
tend to GR when this ratio is small. 

\subsection{Cosmology in the scalar-tensor version of 4D EGB}\label{sec:STversion}

Let us now look at how the equations presented above arise within the 
scalar-tensor version of 4D Einstein-Gauss-Bonnet given by the action \eqref{eq:4Daction1}.
Solving this theory with the line-element \eqref{eq:FRW}, one finds that the scalar 
field equation is given by
\begin{equation}
    \alpha(k+a^2( H+\dot\phi)^2)(\ddot\phi+\dot H+H(\dot\phi+H))=0\,,
\end{equation}
which is solved by
\begin{equation}
\dot \phi = -H + \frac{K}{a}\,,
\label{eq:profile}
\end{equation}
where $K$ is a constant. 

In this formulation of the theory, the Friedmann equation is 
\begin{equation}
H^2+\frac{k}{a^2}= \frac{8\pi G }{3} \rho +\alpha\dot\phi(2 H+\dot\phi)\left(2 \left(H^2+\frac {k}{a^2}\right)+2H\dot \phi+\dot\phi^2\right)\,,
\label{eq:Friedscalar0}
\end{equation}
which, after substitution of the solution given in Eq.~\eqref{eq:profile}, results in
\begin{equation}
H^2+\frac{k}{a^2}+\alpha\left(H^2-\frac{K^2}{a^2}\right)\left(H^2+\frac{K^2+2k}{a^2}\right) = \frac{8\pi G }{3} \rho \,.
\label{eq:Friedscalar}
\end{equation}
It can clearly be seen that this only 
reduces to Eq.~\eqref{eq:Fried1} when $K^2=-k$. For the flat case, this requires $K=0$, and for the 
positively curved case we need $K=\pm i$. Any other value, parametrized by $K^2=-k+C$, where $C$ is a free parameter, leads to an additional dark radiation term, such that the Friedmann equation can be written as 
\begin{equation}
H^2+\frac{k}{a^2}+\alpha \left(H^2+\frac{k}{a^2}\right)^2 = \frac{8\pi G }{3} \rho +\frac{\alpha C^2}{a^4} \,.
\label{eq:FriedDark}
\end{equation}
It is interesting to note that in the positively-curved case a complex scalar field is required to 
set $C=0$ and recover the Friedmann equation of the original theory, and the bouncing behaviour discussed
above. 

Finally, we note that for the perturbed line-element \eqref{eq:FRWpert}, 
taking Eq.~(\ref{eq:profile}) with $K=0$ 
also recovers the perturbed equations \eqref{eq:scalarPerts} of the original theory. In more detail, the first-order equation of motion for the scalar field is, in the flat case,
\begin{equation}
    \alpha K^2\left(\partial^2(\delta\phi+\varphi)+3(\psi''-\delta\phi'')+3K(\psi'+\varphi')\right)=0\,,
\end{equation}
which is automatically solved when $K=0$ and thus does not constrain the field fluctuation, $\delta\phi$. The first-order field equations are modified by the presence of a non-zero $C$ and become
\begin{align}
&8 \pi G a^4 \rho \delta + 2 {\cal A} (- \partial^2 \psi + 3 {\cal H}^2 \varphi + 3 {\cal H} \psi') = 4\alpha K^2\left(3K^2\varphi+3K(\psi'-\delta\phi')+\partial^2(\psi-\delta\phi)\right)\,, \nonumber \\
&{\cal A} \varphi + ({\cal B} - 4 \alpha {\cal H}') \psi =-8\pi G a^4\Pi+2\alpha K^2(\varphi+\psi)\,,\nonumber\\
&{\cal A}\partial^2F_i'+16\pi G(1+w)a^4\rho v_i =2\alpha K^2 \partial^2F_i'\,,\\
&{\cal A}\gamma''_{ij}+2{\cal H} (a^2+2\alpha {\cal H}') \gamma'_{ij}+({\cal B}-4\alpha {\cal H}')\partial^2 \gamma_{ij}=8\pi G a^4\Pi_{ij}+2\alpha K^2(\gamma_{ij}''+\partial^2\gamma_{ij})\nonumber\,,
\label{eq:scalarPerts}
\end{align}
where we have placed the $K$-dependent terms on the right-hand side, to clearly demonstrate that these equations reduce to those of the original theory when $K=0$. Thus, the fact that, in that case, at the linear level, the field perturbation is undetermined is inconsequential, since it affects nothing else. In the general case, in which $K\neq0$, the additional scalar degree of freedom is important and needs to be taken into account for a full description of the solutions, in addition to the extra terms depending on $K$. At the linear level, the additional field perturbations affect only the scalar equations, effectively sourcing the gravitational field. A clear effect of the $K$-dependent terms is on the propagation of gravitational waves, whose speed is now modified, since the parameter $\Gamma$ is then given by $\Gamma \equiv 1 + 2 \alpha (H^2-K^2/a^2)$.

\subsection{Cosmology in diffeomorphism-breaking version of 4D EGB}

The version of 4D Einstein-Gauss-Bonnet that breaks temporal diffeomorphisms differs from the original version by the introduction of a counter-term dependent on the Weyl tensor of the spatial sections, $C_{ijkl}$, and on the Weyl part of a combination of extrinsic curvature tensors, $K_{i[j}K_{k]l}$~\cite{Aoki:2020lig,Aoki:2020iwm}. For that reason, if those Weyl components vanish, the solutions of this version of the theory are the same as those for the original version.

Due to it having conformally-flat spatial sections, the flat FRW space-time is a solution of this version of the theory obeying the same Friedmann equations as the original version of the theory (Eq. (\ref{eq:FRW}), with $k=0$). Perturbations to FRW generally break conformal flatness, but Ref.~\cite{Aoki:2020iwm} showed that this does not happen for scalar fluctuations, since they can always be written as a perturbation to the conformal factor of the spatial metric. Vector and tensor perturbations, however, are expected to introduce a non-zero spatial Weyl tensor, and should obey different equations from those of the original theory. While vectors were not studied yet, the evolution of tensor modes was derived in Ref.~\cite{Aoki:2020iwm}, and obeys
\begin{equation}
\ddot{\gamma}_{ij}+\cbr{3+\frac{4\alpha \dot H}{1+2\alpha H^2}}H \dot{\gamma}_{ij}-c_T^2\frac{\partial^2 \gamma_{ij}}{a^2}+\frac{4 \alpha}{\Gamma a^4}\partial^4 \gamma_{ij}=0\,.
\label{eq:GWsdiff}
\end{equation}
The additional term here has 4 spatial derivatives, and therefore modifies the dispersion relation of gravitational waves, adding a $k^4$ term and substantially modifying the predictions of the theory.

In addition to this, Ref.~\cite{Aoki:2020ila} has studied gravitational waves at second-order in perturbations, as applied to the calculation of the bispectrum of tensor modes from inflation. We do not go into detail here on that scenario, but it is clear from those results that at second-order there are also contributions with 4 spatial derivatives, which would not be present in the original or scalar-tensor versions of the theory.

Next, we review the constraints on 4D Einstein-Gauss-Bonnet from cosmology. In most cases, the constraints apply only to the original theory or to the equivalent scalar-tensor theory with $K=0$. However, when work has been developed in that direction, we also include constraints taking other versions of the theory into account.

\subsection{Constraints on $\alpha$} 

In principle, the background evolution, and the evolution of 
perturbations encoded in Eqs.~\eqref{eq:Fried1}-\eqref{eq:Fried3} and 
Eqs.~\eqref{eq:GWs}-\eqref{eq:scalarPerts}, can be used to constrain 4D Einstein-Gauss-Bonnet, by comparing 
predictions of the theory  against observation, thereby putting constraints on allowed values of $\alpha$. 

Given that $H \approx 2.4 \times 10^{-18} s^{-1}$ 
today, in the late universe a very large value of $\alpha$ is 
required to make $\alpha H^2$ significant. For correction terms to be of the 
same order as terms that arise in GR, for this value of $H$, for example, would require $\alpha \sim 10^{52}m^2$. 
This in turn implies that we expect stronger 
constraints on $\alpha$ to be available when we observe the earlier stages of the 
universe's evolution.

\subsubsection{Gravitational waves. }
First we consider constraints from the propagation of 
gravitational waves. 
The recently detected 
gravitational wave signal GW170817, together with its electromagnetic counterpart, indicate that the 
deviation in the speed of gravitational waves from that of 
light must be less than one part in $10^{15}$ \cite{TheLIGOScientific:2017qsa,Goldstein:2017mmi, Savchenko:2017ffs, Baker:2017hug, Creminelli:2017sry, Ezquiaga:2017ekz, extra}.  From Eq.~\eqref{eq:tensorf} this requires
\begin{align}
\dot{\Gamma}/(H \Gamma) &<10^{-15} \nonumber
\end{align}
or, equivalently,
\begin{align}
\frac{-8 \alpha \epsilon H^2}{c^2 \Gamma} &<10^{-15} \nonumber 
\end{align}
which implies
\begin{align}
|\alpha|  &\lesssim  10^{36} m^2\,,
\end{align}
where 
$\epsilon \equiv -\dot{H}/H^2 $, and 
we have taken $\dot{H}\approx H^2 \approx 5.8 \times 10^{-36} \, {\rm s}^{-2}$, which agrees with estimates 
in Refs.~\cite{Aoki:2020iwm,Feng:2020duo}. The weakness of this constraint is expected given 
that is is a late-time observation.

Ref.~\cite{Feng:2020duo} also produces constraints for the more general scalar-tensor version of 4D Einstein-Gauss-Bonnet with $K\neq0$ described in Section \ref{sec:STversion}. There they conclude that constraints on $\alpha$ are loosened by up to an order of magnitude for particular values of $|K|\sim H_0/2$, and they place a constraint on $K$ of $|K|\lesssim 1.35 H_0$ (see their Fig.~1 in Ref.~\cite{Feng:2020duo} for the detailed region of $\alpha-K$ parameter space excluded, where they use variables $\tilde\alpha=\alpha H_0^2$ and $\tilde A=K/H_0$).

As shown above, the temporal diffeomorphism-breaking version of the theory has a modified equation for gravitational waves with an additional $k^4$ contribution to the dispersion relation. For that reason, the constraints that can be obtained there are very different from those for the original theory. Ref.~\cite{Aoki:2020iwm} showed that the constraint from the speed of gravitational waves is $\alpha\lesssim 10^{-10}\,m^2$, which is the strongest constraint for that version of the theory from any observation.

\subsubsection{From the CMB to  today. }
Next we consider the consistency of the 
equations 
with cosmological observations that inform us directly about the 
expansion history of the universe, 
and about the evolution of structure. Such observations include 
supernovae, and observations of 
the Cosmic Microwave Background (CMB) and of Large Scale Structure (LSS).
This has 
been attempted at varying degrees of sophistication 
by a number of authors\cite{Feng:2020duo,Haghani:2020ynl,Wang:2021kuw}. All 
agree that the theory is consistent with observations so long as 
$\alpha$ is sufficiently small, such that the evolution of the background and 
perturbations is observationally indistinguishable from that in GR. 
The most complete analysis is contained in Ref.~\cite{Wang:2021kuw}, 
which uses CMB, baryon acoustic oscillations, supernovae, and redshift space distortion observations to 
give the constraint  $|\alpha| \lesssim  10^{36} m^2$. The authors assume 
a flat cosmology,  
modify the open-source Boltzman code CAMB 
with the background and perturbed equations for scalar perturbations presented above, and
compare and constrain the parameters of a $\Lambda$CDM-type 
cosmological model together with $\alpha$ using the Monte Carlo Markov chain approach. 
The use of CMB data is key in their study, since it 
probes the universe at energies much higher that those today 
when $\rho/(G \alpha)$ was much larger, and leads to a constraint 
that is orders of magnitude higher than that which can be obtained 
from large scale structure, or supernovae data alone which probe the recent universe.

\subsubsection{Nucleosynthesis. }
The epoch of primordial nucleosynthesis is the earliest 
phase of evolution in the universe for which we have direct observational 
evidence.  The products of this period lead to the abundances of 
the light elements observed in the universe today, and these products are 
sensitive to the expansion of the universe during this phase 
\cite{Carroll:2001bv, Arbey:2011nf}. The energy scale at which 
nucleosynthesis begins is about $1$ MeV, corresponding to  $H^2\approx 0.14 \, {\rm s}^{-2}$, and 
requiring that the  $\alpha H^2$ is less than unity at this time implies that  $\alpha <  10^{18} \, {\rm m}^2$. 
A more precise 
bound comes from simulating this epoch utilising Eq.~\eqref{eq:Fried4}. In Ref~\cite{Clifton:2020xhc} 
the open source code PRIMAT\footnote{http://www2.iap.fr/users/pitrou/primat.htm} \cite{Pitrou:2018cgg} was used 
for this purpose leading to the bound $|\alpha| \lesssim  10^{17} \, {\rm m}^2$. 

\subsubsection{Inflation. }
At earlier times than primordial nucleosynthesis 
the evolution of the universe is less well understood. Nevertheless, 
the very early Universe is believed to have undergone a 
period of accelerated 
expansion, known as ``inflation'' that 
is the origin of structure in the universe. The behaviour of this phase of 
evolution within 4D Einstein-Gauss-Bonnet was studied in Ref.~\cite{Clifton:2020xhc} and previously 
in Ref.~\cite{Casalino:2020pyv}, and we now summarise the results.

Inflation leads to a flat universe, and as 
noted above for a flat universe, there is an upper 
limit on $H$ given by $H^2 = 1/(-2 \alpha) $ when $\alpha$ is negative. 
Assuming that inflation takes place above the TeV scale, which is consistent with the lack of new physics at the LHC and also the need for Baryogenesis, this implies that $ \alpha \gtrsim  -10^{-6} \, {\rm m}^2$, which is a very tight constraint on negative $\alpha$.

Unfortunately, less can be said in the case of positive $\alpha$. Assuming 
inflation to be driven by a single canonical scalar field $\chi$ with potential energy $V(\chi)$, Ref.~\cite{Clifton:2020xhc} found that 
the condition for inflation $\epsilon \equiv -\dot{H}/H^2 < 1$
translated to the requirement 
\begin{equation}
\frac{2 V'^2 \alpha^2}{9 m_{\rm pl}^2 \Gamma_V(\Gamma_V-1)^2} \lesssim 1 \, ,
\end{equation}
where  $\displaystyle \Gamma_V = \sqrt{1 + 4 V\alpha/(3 m_{\rm pl}^2})$, and where here 
a dash indicates a derivative with respect to $\chi$. This implies 
that the slope of the potential (i.e. $V'/V$) must be shallower in 4D Einstein-Gauss-Bonnet
than in GR, but that even when the correction terms in the Friedmann equation 
dominate (when $\alpha H^2 \gg1$), we can always assume a sufficiently flat form of $V(\chi)$ in order 
to allow inflation to still proceed.

Turning to perturbations, inflation produces both scalar fluctuations and gravitational waves. During inflation 
the scalar fluctuation most commonly studied is the 
uniform density curvature perturbation, an action for which is given by
\begin{equation}
\label{eq:zetaf}
S_\zeta= \frac{1}{2}\int {\rm d}^3 x {\rm d} t \,a^3 \Gamma \epsilon \left ( \dot{\zeta}^2  - c^2 \frac{(\partial \zeta )^2}{a^2} \right)\,.
\end{equation}
Eqs.~\eqref{eq:zetaf} and (\ref{eq:tensorf}), imply that as in GR, 
the curvature and tensorial perturbations are conserved on 
super-horizon scales. These equations therefore lead directly to expressions 
for the spectra of tensor and scalar perturbations 
in terms of quantities at the time a given wavenumber, $k$, crosses the apparent horizon, which results in
\begin{equation}
P_T = \left.\frac{2}{\pi^2 m_{\rm pl}^2 }\frac{H^2}{\Gamma}\right |_*  \quad ~ ~ {\rm and} \quad ~ ~ P_\zeta = \left. \frac{1}{8\pi^2 m_{\rm pl}^2} \frac{H^2}{\epsilon \Gamma}\right |_* \, ,
\end{equation}
where the asterisk indicates quantities are to be evaluated at horizon crossing. 

These expressions imply that the tensor-to-scalar ratio takes its usual form $r=16\epsilon$, while differentiating the power spectra with respect to horizon crossing scale $k=aH$ gives the spectral indices at leading-order in slow roll as
\begin{align}
\frac{\partial \log P_\zeta }{\partial \log (a H) } &= n_s -1 =  - 2 \epsilon - \dot \epsilon/(\epsilon H)- \dot \Gamma/(\Gamma H)\,,\nonumber\\
\frac{\partial \log P_T }{\partial \log (a H) } &= n_T = -2 \epsilon- \dot \Gamma/(\Gamma H)\,.
\end{align}
Since $n_T \neq -r/8$, the consistency equation of canonical single field inflation is violated.  The form of $n_s$, while different from its canonical form, can be made in agreement with 
observational constraints on this quantity \cite{Akrami:2018odb}  
through a suitable choice of $V(\chi)$ even when $\alpha H^2 \gg 1$.

In the absence of a detection of gravitational waves from inflation, therefore, there is currently no way of 
distinguishing inflation in 4D Einstein-Gauss-Bonnet from inflation in GR, or placing constraints on positive values of 
$\alpha$ using inflation. It would be interesting to calculate higher-order correlations of the inflationary perturbations, that might also alleviate this degeneracy, but 
as yet such a calculation has not been attempted.

\section{Weak Field and Constraints on $\alpha$}\label{sec:wf}

As with any newly proposed theory of gravity, it is important to understand the weak-field behaviour of 4D Einstein-Gauss-Bonnet. It is the weak-field limit to which we have the most direct access, and in which the vast majority of experimental and observational tests of gravitational theories have so far been performed. Such a limit is sufficient to describe almost all astrophysical systems, with the notable exceptions of neutron stars, black holes and cosmology, which have already been discussed in the preceding sections of this review.

In order to explore the weak-field behaviour of 4D Einstein-Gauss-Bonnet we will deploy the standard framework; a post-Newtonian expansion of the field equations and equations of motion of the theory order-by-order in the typical velocity of bodies in the system as a fraction of the speed of light, $v/c$. Such an expansion is the bedrock of almost all weak-field gravity phenomenology, in the solar system as well as in extra-solar systems such as binary pulsars \cite{will2018theory}. We will also restrict ourselves to the scalar-tensor version on this theory, as given in Eq. (\ref{feqs}).

\subsection{Post-Newtonian expansions}

The first step in performing a post-Newtonian analysis is to expand the metric around Minkowski space, such that the metric of space-time is approximated as
\begin{equation}
g_{\mu \nu} = \eta_{\mu \nu} + h_{\mu \nu} \, ,
\end{equation}
where $h_{\mu \nu}$ is a small perturbation to the metric of Minkowski space, $\eta_{\mu \nu}$. The components of $h_{\mu \nu}$ are then expanded in the order-of-smallness they are in comparison to $v/c$, such that 
\begin{align*}
h_{00} = h^{(2)}_{00} + h^{(4)}_{00} + O\left(\frac{v^5}{c^5}\right) \,, \quad
h_{0i} = h^{(3)}_{0i} + O\left( \frac{v^4}{c^4} \right) \,, \quad {\rm and} \quad 
h_{ij} = h^{(2)}_{ij} +O\left( \frac{v^3}{c^3} \right) \, ,
\end{align*}
where the superscripts in brackets refer to the order-of-magnitude of a particular object in $v/c$ (i.e. such that $h^{(2)}_{00} \sim v^2/c^2$). The corresponding expansion for the mass density and pressure in matter fields are given by
\begin{equation} \nonumber 
\rho = \rho^{(2)}, \quad p = p^{(4)}, \quad {\rm and} \quad \Pi = \Pi^{(2)} \, ,
\end{equation}
where $\Pi$ is the internal energy per unit mass of a body. 

The remaining ingredient of the post-Newtonian expansion is then the requirement that the time-variation of any quantity associated with the geometry or stress-energy in the system does not vary more quickly than the position of the bodies themselves, such that the time-derivative of any quantity is always an order-of-smallness in $v/c$ compared to any spatial derivative of the same quantity, which we can think of schematically as the rule:
\begin{equation}
\frac{\partial \;}{\partial t} \sim \frac{v}{c} \, \frac{ \partial \;}{\partial x} \, .
\end{equation}
In the case of the regularized scalar-tensor versions of 4D Einstein-Gauss-Bonnet, the only remaining object is the extra scalar degree-of-freedom, which can be expanded as
\begin{equation}
\phi = \phi^{(0)} + \phi^{(2)} +  O\left( \frac{v^4}{c^4} \right) \, .
\end{equation}

Using these expansions in the field equations of the regularized scalar-tensor 4D Einstein-Gauss-Bonnet theories, one finds that the leading-order parts of the perturbations to the metric are given by \cite{Clifton:2020xhc}
\begin{equation}
h^{(2)}_{00} = 2 \, U \,, \quad h^{(2)}_{ij}= 2\, U \, \delta_{ij} \, \quad {\rm and} \quad h^{(3)}_{0i} = -\frac{7}{2} V_i-\frac{1}{2} W_i \, ,
\end{equation}
where $U$ is the Newtonian potential that satisfies $\nabla^2 U = -4 \pi \, \rho$, and $V_i$ and $W_i$ are the standard post-Newtonian vector gravitational potentials (see e.g. \cite{will2018theory}). At this point the metric perturbations of the theory are all identical to those obtained from Einstein's theory, but we will now see that this is no longer the case when we go to next-to-leading order in the metric perturbations, and in particular when $h^{(4)}_{00}$ is calculated.

In determining the post-Newtonian perturbation $h^{(4)}_{00}$ it is necessary to determine the perturbation to the scalar field, which can be found to be given by $\phi^{(2)}= \pm U$, where the $\pm$ sign occurs due to the $\phi^{(2)}$ perturbation appearing in quadratic combinations at the required order. Using this information, one can then determine that
\begin{equation} 
h_{00}^{(4)} = -2 U^2 +4 \Phi_1 + 4 \Phi_2 + 2 \Phi_3 + 6 \Phi_4 \mp \left( \frac{\alpha}{4\pi} \right) \Phi_{\mathcal{G}} \, ,
\end{equation}
where $\{ \Phi_1, \Phi_2, \Phi_3, \Phi_4 \}$ are standard post-Newtonian scalar gravitational potentials, and where $\Phi_{\mathcal{G}}$ is given by \cite{Clifton:2020xhc}
\begin{equation}
\Phi_{\mathcal{G}} =  \int \frac{\mathcal{G}^{(4)}}{\vert {\bf x} - {\bf x'} \vert} d^3 x'
=8 \int \frac{U_{,ij} \, U_{, ij} - \left( \nabla^2 U \right)^2  }{\vert {\bf x} - {\bf x'} \vert} d^3 x'  \, ,
\end{equation}
where $\mathcal{G}^{(4)}$ is the Gauss-Bonnet term expanded to order $v^4/c^4$, and where $\alpha$ is the coupling constant of the theory. The reader may note that all post-Newtonian parameters in this theory are identical to those of GR (i.e. $\beta =\gamma = 1$ and $\xi = \alpha_1=\alpha_2=\alpha_3=\zeta_1=\zeta_2=\zeta_3=\zeta_4=0$), but that there exists an extra post-Newtonian potential in $h_{00}^{(4)}$. This new potential takes the form of the gravitational field of the Gauss-Bonnet term $\mathcal{G}$ itself, and is proportional to the coupling constant $\alpha$.

\subsection{Constraints on $\alpha$}

After calculating the relativistic Lagrangian and Hamiltonian of the relevant 2-body dynamics, it can be found that the precession of periapsis in bound orbits in this theory is given by the following expression \cite{Clifton:2020xhc}:
\begin{equation} \label{prec}
{\delta \varphi = \frac{6 \pi M}{ a (1-e^2)} \left[ 1 \pm \frac{\alpha (4+e^2)}{a^2(1-e^2)^2} \right] }\, ,
\end{equation}
where $a$ and $e$ are the semi-major axis and eccentricity of the orbit, and $M$ is the total mass of the two bodies. The contribution of the Gauss-Bonnet term can be seen in the second term in brackets, which is added to the usual general relativistic expression (see e.g. \cite{landau1975classical}). The $\pm$ sign in this expression comes directly from the corresponding sign in the equation for $\phi^{(2)}$, and the extra term can be seen to vanish when $\alpha=0$.

The theory above can be seen to give a non-negligible contribution to the precession of orbits, but the result that $\gamma=1$ in this theory (as it does in GR) means that the deflection of light and the Shapiro time delay of radio signals cannot be used to distinguish 4D Einstein-Gauss-Bonnet from GR. These are usually two of the most discriminating observational effects of relativistic gravity, so it is notable that in this case they do not provide any way to distinguish 4D Einstein-Gauss-Bonnet from Einstein's original theory. Instead, we must use the post-Newtonian gravitational dynamics of time-like particles in order to test the theory. We will summarize the relevant constraints that result below, which are obtained using Eq. (\ref{prec}).

First, observations of the classic perihelion precession of the planet Mercury \cite{pitjeva2013relativistic} can be used to constrain the magnitude of the coupling parameter of the theory to be given by
\begin{equation}
\vert \alpha \vert = \vert (-3.54 \pm 5.31) \vert \times 10^{16} \, {\rm m^2} \, .
\end{equation}
Changing the set of observations used provides different constraints, but it appears that all current observations are at the level $\vert \alpha \vert \lesssim 10^{17} \, {\rm m^2}$. Alternatively, using observations of the artificial LAGEOS satellites \cite{Lucchesi:2010zzb} gives the considerably more stringent bound
\begin{equation}
\vert \alpha \vert = \vert (0.23 \pm 1.74) \vert \times 10^{10} \, {\rm m^2}  \, .
\end{equation}
Looking further afield, 
the double pulsar PSR J0737-3039A/B \cite{lyne2004double} allows observations that can be used to infer \cite{Kramer:2006nb}
\begin{equation}
\vert \alpha \vert = \vert (0.4 \pm 2.4) \vert \times 10^{15} \, \frac{\rm m^2}{\sin i} \, .
\end{equation}
where $i$ is the angle of inclination of the system. Further discussion of these results, can be found in Ref. \cite{Clifton:2020xhc}.

Finally we can consider the ``table-top'' tests of gravity, which are laboratory tests of gravity that can be used to test gravitational potentials of the form 
\begin{align}
V =  &\frac{m_1 m_2}{r} \left(1 - \frac{A_n}{ r^n} \right)\,,
\end{align}
where $A_n$ is a set of constants, where $m_1$ and $m_2$ are the masses of the bodies involved, and where no sum is implied over $n$. Comparison with the weak-field solutions presented above shows that the 4D Einstein-Gauss-Bonnet theories can be understood within this framework if we take $A_3 = \pm 2\alpha \,  { (m_1+m_2)}$ (with all other $A_n$ vanishing). Observational constraints from experiment \cite{PhysRevLett.116.131101} then yield the bound \cite{Clifton:2020xhc}
\begin{equation}
\vert \alpha \vert \lesssim 10^{16}\, {\rm m}^2 \, .
\end{equation}
This result can be seen to be competitive with the bounds obtained from observations of the orbit of Mercury and the orbits of the double pulsar, but are weaker than those obtained from the LAGEOS satellites.

\par Including recent gravitational wave observations, tighter constraints can be imposed. Recall that a black hole in the 4DEGB theory has horizons given by Eq. \eqref{eq:horizons}, such that $r_+$ should be a real value, imposing that for any observed black hole we should have $M\geq \sqrt{\alpha}$. This reality condition imposes tight constraints on $\alpha$, given that the theory should be able to reproduce the lightest observed black holes. Upper bounds on $\alpha$ obtained with this method are given in Table \ref{tab:GWconstraints} (see Refs. \cite{Clifton:2020xhc,Charmousis:2021npl} for details)\footnote{These constraints assume all compact objects involved to be black holes. Note that estimates for the mass of the bodies rely on the extraction of the chirp mass and the mass ratio in the inspiral phase assuming GR.}.

\begin{table}
\begin{center}
\begin{tabular}{||c | c | c ||} 
 \hline
 Event & Upper bound on $\alpha$ (m$^2$) & Data from Ref. \\ [0.5ex] 
 \hline\hline
 GW150914 & $\sim 10^9$ & \cite{LIGOScientific:2016wyt}\\ 
 \hline
 GW170608 & $\sim 10^8$ & \cite{LIGOScientific:2017vox}\\
 \hline
 GW190814 & $\sim 10^7$ & \cite{LIGOScientific:2020zkf}\\
 \hline
 GW200115 & $\sim 10^8$ & \cite{LIGOScientific:2021qlt}\\
 \hline
\end{tabular}
\end{center}
\caption{Upped bounds on $\alpha$ from gravitational wave events, by imposing the reality condition on the horizon of the black holes.}
\label{tab:GWconstraints}
\end{table}

\par Regarding negative values of $\alpha$, Ref. \cite{Charmousis:2021npl} imposes a very tight constraint
\begin{equation}
    -\alpha \lesssim 10^{-30} \mbox{m}^2,
\end{equation}
by requiring that atomic nuclei should not be shielded by a horizon. For all practical purposes, negative values of the coupling can be excluded from future analysis as the associated gravitational effects would be undetectable.

\section{Summary}

The 4D Einstein-Gauss-Bonnet theories of gravity have sparked a very considerable amount of interest over the past couple of years, and this review is aimed at providing a balanced overview of that work. We started the review with an introduction to Gauss-Bonnet terms in the action for gravity, in the context of the work of Lovelock and Lanczos. Such terms are well-motivated from a theoretical perspective, and have appeared in very many different approaches to generalizing Einstein's theory, as well as from more fundamental theories such as string theory.

The Gauss-Bonnet term in the gravitational action is usually taken to have no consequences for the field equations of gravity in 4 space-time dimensions. This result, which has been well-known since the early days of general relativity, was recently called into question by Glavan \& Lin \cite{Glavan:2019inb}. These authors suggested that it may be possible for the Gauss-Bonnet term to have non-trivial consequences even in $D=4$, if the coupling constant of this term is taken to diverge in a suitable way as $D \rightarrow 4$, in a manner reminiscent of the dimensional regularization approach used in quantum field theory. The novelty of this idea, the simplicity of the space-times that result from it, and the controversies associated with it, are what has fuelled the recent interest in this area, and what we have aimed to capture in this review.

The technical details of the Glavan \& Lin's proposal are explained in Section \ref{sec:4D EGB} of this review, along with the criticisms that followed from it. In particular, we have sought to explain why their proposal does {\it not} evade Lovelock's theorem in the way that was envisaged in their original work, by providing citations to the primary literature on this point, and by reproducing the essence of these arguments. This is followed by a selection of the proposed theories that provide a well-motivated alternative to the original scheme of Glavan \& Lin, and which share some of the features of their approach, including the intriguing solutions that they found for the geometry of space-time. All of these approaches violate the assumptions of Lovelock's theorem, by either including extra fundamental fields, higher dimensions, or by breaking diffeomorphism invariance of the theory. Nevertheless, they provide an interesting new class of alternative theories of gravity, which themselves provide the primary focus of the remainder of our review.

Subsequently, in Sections \ref{sec:bh}, \ref{sec:cos} and \ref{sec:wf} we explore the solutions of these theories as relevant for the study of black holes, cosmology, and the weak-field limit of gravity, respectively. A large number of interesting phenomena have been discovered in these different areas of gravitational physics, and we aim to give an overview of some of that work, as well as citations to the primary literature. In particular, we discuss the existence of exact solutions, and their properties. This includes a discussion of the properties of singularities and the thermodynamics of black hole horizons in these theories.

The cosmologies of these theories, as discussed in Section \ref{sec:cos}, also demonstrate an interesting set of behaviours. In particular, it can be seen that different branches exist in the solutions to the Friedmann equations, and that some ranges of values of the coupling constant are restricted if we wish to produce viable cosmological models. In the weak-field limit of these theories, discussed in Section \ref{sec:wf}, we find that a lot of the gravitational phenomena that results from them is observationally indistinguiashable from general relativity, particularly with regards to the lensing and time-delay of light, which usually provides amongst the tightest constraints on relativistic gravity. Instead, we find that the LAGEOS satellites provide the tightest bounds on the coupling parameter of these theories.

Overall, the current observational bounds on the coupling parameter of this theory indicate that
\begin{equation}
    0< \alpha < 10^{10} \, m^2 \, ,
\end{equation}
where the lower bound comes from early universe cosmology and atomic nuclei, and the upper bound comes from constraints from the LAGEOS satellites. Indicative results from preliminary calculations show that these constraints could be even tighter if we include recent gravitational wave observations, reducing to 
\begin{equation}
    0< \alpha \lesssim 10^{7} \, m^2 \, ,
\end{equation}
Such tight constraints would mean that deviations from general relativity, of the type given by the 4D Einstein-Gauss-Bonnet theories, would be restricted to being large only in the very early universe or in the immediate vicinity of black holes.

Looking forward, there is still plenty of work that needs to be done on these theories in order to fully understand them. In particular, numerical simulations of the merger of black holes, and the associated gravitational radiation emitted, is currently lacking. It is expected that this would provide the best way to constrain this theory, and so it would be extremely interesting to see such simulations performed. On the mathematical side, the initial value problem of these theories has not yet been proven to be well posed (despite these theories belonging to the Horndeski class, and proofs of well-posedness existing for a sub-set of these \cite{Papallo:2017qvl}). It also appears that the sub-set of regularized theories without a canonical kinetic term for the scalar might exhibit a strong coupling issue, but a more careful analysis is also required on this point. Finally, there are no exact rotating black hole solutions known to these theories, despite the simplicity of the spherically symmetric vacuum and electrovacuum cases. We also note that new approaches to 4D EGB are also still being developed, such as the intriguing study in Ref. \cite{Sengupta:2021mpf}, which considers an extra dimension of vanishing proper length \cite{Sengupta:2019ydf}. All of these areas, and more, remain to be fully studied in order to have a complete understanding of this interesting collection of theories. We look forward to such developments, and to seeing where the study of 4D Einstein-Gauss-Bonnet gravity will be taken next.

\newpage

\section*{Acknowledgements}

PF is supported by the Royal Society grant RGF/EA/180022 and acknowledges support from the project
CERN/FISPAR/0027/2019. DJM is supported by a Royal Society University Research Fellowship. TC acknowledges financial support from the STFC under grant ST/P000592/1. PC acknowledges support from a UK Research and Innovation Future Leaders Fellowship (MR/S016066/1).

\section*{Bibliography}


\end{document}